\documentclass[journal,draftcls,onecolumn,12pt,twoside]{IEEEtran}
\normalsize
\usepackage{amsmath,amssymb,amsfonts,amsthm}
\usepackage[square, comma, sort&compress, numbers]{natbib}
\usepackage{algorithm, algpseudocode}
\usepackage{array}
\usepackage{multicol}
\usepackage{multirow,booktabs}
\usepackage{graphicx}
\usepackage{graphicx,pgf,tikz}
\usepackage{setspace}
\usepackage{bbm}
\usepackage{dsfont}
\usepackage[perpage]{footmisc }
\usetikzlibrary{arrows,automata}

\usepackage{subfigure}

\theoremstyle{definition}
\newtheorem{theorem}{Theorem}
\newtheorem{lemma}{Lemma}

\allowdisplaybreaks

\makeatletter
\renewcommand\normalsize{
	\abovedisplayskip 3\p@ \@plus5\p@ \@minus7\p@
	\belowdisplayskip \abovedisplayskip
	\let\@listi\@listI}
\makeatother

\begin{document}
\IEEEoverridecommandlockouts
\title{Delay-Optimal and Energy-Efficient Communications with Markovian Arrivals}

\author{
Xiaoyu~Zhao,
Wei~Chen,~\IEEEmembership{Senior~Member,~IEEE,}
Joohyun~Lee,~\IEEEmembership{Member,~IEEE,}
and~Ness~B.~Shroff,~\IEEEmembership{Fellow,~IEEE}
\thanks{
X. Zhao and W. Chen are with the Department of Electronic Engineering and Beijing National Research Center for Information Science and Technology, Tsinghua University. E-mail: xy-zhao16@mails.tsinghua.edu.cn, wchen@tsinghua.edu.cn.

J. Lee is with the Division of Electrical Engineering, Hanyang University. E-mail: joohyunlee@hanyang.ac.kr.

Ness B. Shroff holds a joint appointment in both the Department of ECE and the Department of CSE at The Ohio State University. E-mail: shroff.11@osu.edu.

This research was supported in part by the National Natural Science Foundation of China under Grant No. 61671269, the Beijing Natural Science Foundation under Grant No. 4191001, and the National Program for Special Support for Eminent Professionals of China (10,000-Talent Program).
}}

\maketitle

\begin{abstract}

In this paper, delay-optimal and energy efficient communication is studied for a single link under Markov random arrivals.
We present the optimal tradeoff between delay and power over Additive White Gaussian Noise (AWGN) channels and extend the optimal tradeoff for block fading channels.
Under time-correlated traffic arrivals, we develop a cross-layer solution that jointly considers the arrival rate, the queue length, and the channel state in order to minimize the average delay subject to a power constraint.
For this purpose, we formulate the average delay and power problem as a Constrained Markov Decision Process (CMDP).
Based on steady-state analysis for the CMDP, a Linear Programming (LP) problem is formulated to obtain the optimal delay-power tradeoff.
We further show the optimal transmission strategy using a Lagrangian relaxation technique.
Specifically, the optimal adaptive transmission is shown to have a threshold type of structure, where the thresholds on the queue length are presented for different transmission rates under the given arrival rates and channel states.
By exploiting the result, we develop a threshold-based algorithm to efficiently obtain the optimal delay-power tradeoff.
We show how a trajectory-sampling version of the proposed algorithm can be developed without prior need of arrival statistics.

\end{abstract}


\IEEEpeerreviewmaketitle

\newpage

\section{Introduction}

There is increasing interest in developing strategies to achieve low-latency transmissions in a wide variety of applications, e.g., in mission critical applications for the Internet of Things (IoT), or Ultra Reliable and Low Latency Communications (URLLC) in Fifth-Generation (5G) systems \cite{URLLC_req,URLLC_req_3}. At the same time, there is also a push towards developing strategies to make devices and networks more energy efficient \cite{Power_efficient_1,Power_efficient_2}. Thus, in our work, we will aim to understand the fundamental tradeoff between delay and energy. More specifically, we will develop a cross-layer solution that minimizes the delay for a given power constraint.
 
\vspace{3mm}

Cross-layer design has been used as a potential enabler to satisfy the requirements of low latency \cite{URLLC_allocation}.
In \cite{cross_layer_delay_1}, a tradeoff between delay and throughput was established based on a cross-layer design that combines adaptive modulation and coding with a truncated Automatic Repeat reQuest (ARQ).
In \cite{djonin2007mimo}, the authors proposed a cross-layer power and rate allocation control to minimize power consumption with a delay constraint in Multiple-Input Multiple-Output (MIMO).
Moreover, energy-efficient cross-layer designs are also studied for packet transmission in wireless networks.
In \cite{power_efficient}, a cross-layer online algorithm was proposed to obtain a more energy efficient transmission over wireless networks.
For multi-hop wireless networks, a cross-layer framework was also presented to jointly consider power control and scheduling in \cite{power_ef_cross_layer_2}.
With the stringent requirements in 5G, the cross-layer designs have been studied to achieve the low latency and energy efficient transmissions in multiple scenarios, such as tactile Internet \cite{URLLC_cross_layer_3} and wireless mesh network \cite{power_ef_cross_layer}.

\vspace{3mm}

In this work, we take a cross-layer design approach to analytically establish the power-delay tradeoff.
To jointly optimize the delay and power, the design problem can be formulated using a Markov Decision Process (MDP).
In \cite{collins1999transmission}, Collins and Cruz considered cross-layer scheduling of an adaptive transmitter over a two-state fading channel.
In their work, the authors established a tradeoff between the average delay and power consumption based on Dynamic Programming (DP), where the only objective of the MDP is formulated as the weighted sum of average power and delay.
Follow-up papers \cite{berry2002communication, berry2013optimal, rajan2004delay} extended this study in various directions with the DP formulation in \cite{collins1999transmission} employed.
In \cite{berry2002communication}, Berry and Gallager formulated the optimal delay-power tradeoff curve for a multi-state block fading channel, where the fixed-length coding and variable-length coding are discussed.
With the DP formulation, the authors have presented all the Pareto optimal power-delay operating points and studied the optimal tradeoff in the regime of asymptotically large delays.
For the regime of asymptotically small delays, Berry has further presented the behavior of the optimal delay-power tradeoff in \cite{berry2013optimal}.
Moreover, a single-parameter scheduler, labeled log-linear scheduler, was proposed over a block fading channel in \cite{rajan2004delay} with near-optimal performance.
In our previous work \cite{chen2007optimal}, the optimal delay-power tradeoff was attained by formulating a Constrained MDP.
With a probabilistic scheduling framework employed, we converted the CDMP problem as an LP problem.
By solving the derived LP problem, we obtain an arbitrary power-delay operating point on the optimal tradeoff curve.

\vspace{3mm}

We further focus on the structural properties of the optimal transmission policies in the cross-layer design.
By exploiting structural properties of the optimal policy, a substantial reduction in computational complexity can be obtained for finding the optimal delay-power tradeoff.
For example, in \cite{goyal2003power}, the structure of the optimal policy were investigated for an adaptive transmitter over the fading channel with interference.
The authors of \cite{ata2005dynamic} further developed an explicit formula for the optimal transmission rate, through which the optimal rate of the single link over a static channel is expressed as a increasing function of queue length.
In \cite{ngo2010monotonicity}, the optimal scheduling was presented in correlated fading channel with the ARQ protocol employed.
The monotonicity of the optimal scheduling was also shown by presenting the optimal rate as an increasing function of the buffer occupancy.
Moreover, by using the policy structures, the complexity of point-to-point network transmission control in \cite{MDP_communication_1} was effectively reduced with the tools from graph signal processing employed for large state space.
In \cite{MDP_communication_2}, based on the structural properties, a novel accelerated reinforcement learning (RL) algorithm was formulated for an energy-harvesting wireless sensor with latency-sensitive data.
Based on the formulated LP problem in our previous work \cite{chen2007optimal}, we also shown a threshold-based structure for the optimal transmission policies.
For the optimal threshold-based policy, we further give a detailed description by showing that the transmission rates are selected deterministically for all the queue lengths except a particular threshold.
The work about the optimal threshold-based policies was also extended to the communication systems with adaptive transmission \cite{chen2017delaytcom}, arbitrary burstiness random arrival \cite{MwangTCOM}, and multi-state fading channels \cite{ARQ}, respectively.

In this work, we generalize our previous work in \cite{xiang_ICC_2017} to show delay optimality with Markov arrivals.
Our generalization is motivated by the work of \cite{self_similar_2}, where network arrivals are shown to exhibit time-correlations.
By modeling the user's arrival as a Markov chain, we first present a cross-layer design to determine the transmission rate.
In particular, we determine the transmission rates by its probability distribution, which is obtained for the current queue length, arrival rate, and the channel state.
With the degenerated probability distribution employed, we can present a deterministic rate selection as the special case for probabilistic transmission policies.
Under the probabilistic cross-layer design, we then formulate the adaptive transmission as a CMDP.
In this way, we next show delay optimality for AWGN channels, where the impacts of the Markovian arrivals is presented for the optimal delay-power tradeoff.
Furthermore, the optimal tradeoff between the delay and power consumption is extended to block fading channels.

For AWGN channels, we first convert the formulated CMDP as an equivalent LP problem.
By this means, we construct the optimal delay-power tradeoff to minimize the average delay under an average power constraint.
We further show the optimal tradeoff by using a curve that consists of all the optimal power-delay pairs for different power constraints.
We refer the curve as the optimal delay-power tradeoff curve, and show the typical geometric properties of it under Markovian arrivals, i.e., the tradeoff curve is piecewise linear, decreasing, and convex.
By jointly exploiting the properties of both the optimal tradeoff curve and the corresponding optimal policies, we then show that the optimal average delay is generated by a threshold-based optimal adaptive transmission policy.
Based on the threshold-based structure, we finally develop an algorithm to efficiently determine the optimal transmission strategies, through which the optimal delay-power tradeoff is presented.
In practice, we show that an online version of the threshold-based algorithm can be also exploited without any need for random arrival statistics.

\vspace{3mm}

Moreover, we extend the optimal delay-power tradeoff by considering block fading channels.
With a block fading channel employed, we can obtain the equivalent LP problem for the adaptive transmitter that is derived based on the formulated CMDP.
We then obtain a similar threshold-based structure on the queue length for fading channels.
As a result, with the current arrival rate and channel state given, we can particularly attain the corresponding transmission rate by comparing the current queue length with the thresholds for different transmission rates.

\vspace{3mm}

The rest of this paper is organized as follows. In Section II, the system model is presented as a CMDP.
By formulating the CMDP as an LP problem, Section III investigates the optimal delay-power tradeoff over AWGN channels.
Then, the corresponding optimal transmission policy is presented in Section IV under the threshold-based structure.
In Section V, we further extent the optimal delay-power tradeoff over a block fading channel.
Finally, numerical results and conclusions are given in Sections VI and VII, respectively.

\begin{figure}[t]
	\centering
	\includegraphics[width=1\columnwidth,height=0.25\columnwidth]{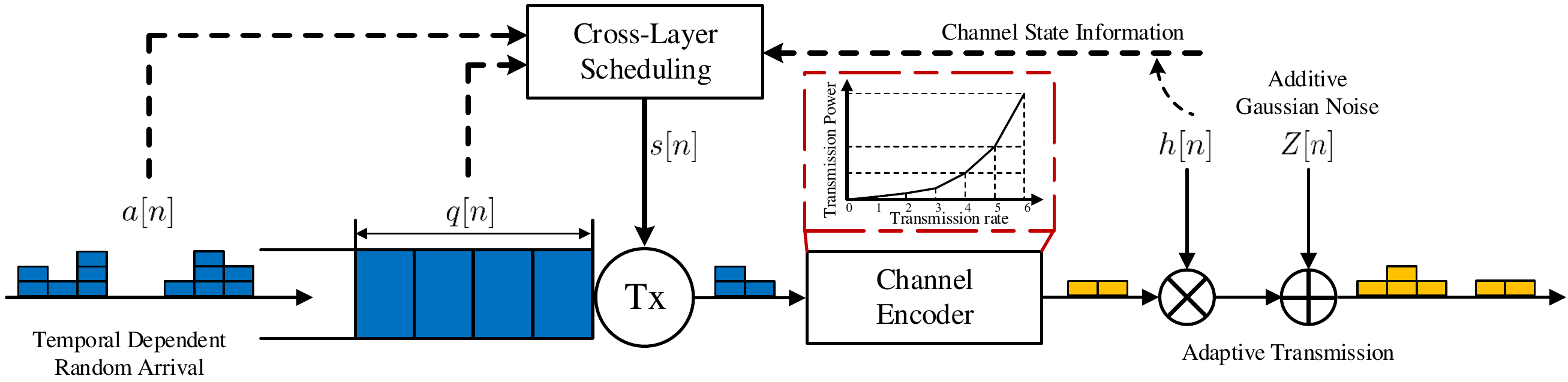}%
	\vspace{-8mm}
	\caption{System Model}
	\vspace{-4mm}
	\label{fig_model}
\end{figure}

\section{System Model}
\vspace{-2.5mm}

In this paper, we focus on a single link of an adaptive transmitter that serves traffic arriving according to a general Markovian process.
As shown in \figurename~{\ref{fig_model}}, the system is assumed to be time-slotted.
The data packets arrive at the beginning of each timeslot according to a stationary and ergodic Markov chain that has finite states.
The state of the Markov chain corresponds to the number of packets that arrive in timeslot $n$, and is denoted by $a[n]$, where the maximum value of $a[n]$ is defined as $A$, i.e., $a[n]\in\{0,1,\cdots,A\}$.
Given that $a[n]$ packets arrive in timeslot $n$, $a[n+1]$ is characterized by the transition probability $\gamma_{a,a'}$ that is \mbox{defined as}%
{\setlength\abovedisplayskip {2pt plus 7pt minus 8pt}
	\setlength\belowdisplayskip {2pt plus 7pt minus 8pt}
\vspace{-2mm}
\begin{equation}
\gamma_{a,a'}=\Pr\{a[n+1]=a'~|~a[n]=a\},
\label{transition_arrival}
\vspace{-2mm}
\end{equation}}%
where $a$ and $a'$ belong to set $\{0,1,\cdots,A\}$.
In other words, the probability that $a[n+1]=a'$ is shown as $\gamma_{a,a'}$ given that $a[n]=a$.
Note that $\gamma_{a,a'}\ge{}0$ and $\sum_{a'=0}^{A}\gamma_{a,a'}=1$.
With the transition probabilities $\gamma_{a,a'}$, $\alpha$, the expected number of arrivals in a timeslot, is given by
{\setlength\abovedisplayskip {2pt plus 7pt minus 8pt}
	\setlength\belowdisplayskip {2pt plus 7pt minus 8pt}
\vspace{-2mm}
\begin{equation}
\alpha=\sum_{a=0}^A a\phi_a,
\vspace{-1mm}
\end{equation}}%
where $\phi_a$ denotes the steady-state probability of $a$ arrivals in a timeslot.

Arriving packets enter a buffer of size $Q$.
At each time $n$, the queue length $q[n]$ belongs to set $\{0,1,\cdots,Q\}$, and evolves as 
{\setlength\abovedisplayskip {2pt plus 7pt minus 8pt}
	\setlength\belowdisplayskip {2pt plus 7pt minus 8pt}
\vspace{-2mm}
\begin{equation}
q[n+1]=\min\{\max\{q[n]-s[n],0\}+a[n+1],Q\},
\label{queue_update}
\vspace{-2mm}
\end{equation}}%
where $s[n]$ denotes the number of packets that are transmitted in timeslot $n$.



Due to the limited throughput at the transmitter, the number of packets that can be transmitted in each timeslot is upper bounded by $S$.
The transmission rate $s[n]$ belongs to the set $\{0,1,\cdots,S\}$.
We then assume that the maximum transmission rate is greater than or equal to the maximum data arrival rate, i.e., $S\ge{}A$.
As a result, we provide the stability of the queue system under an arbitrary Markov arrival process, where the average arrival rates can range from $0$ to $A$ under different arrival processes.
Further, to avoid underflow and overflow of the buffer, $s[n]$ needs to satisfy $0\le{}q[n]-s[n]\le{}Q-A$.
In other words, for each given queue length $q$, we have $q-Q+A\le{}s\le{}q$.
Therefore, with a given queue length $q$, we define the feasible region $\mathcal{S}(q)$ of the transmission rate as $\{s|\max\{q-Q+A,0\}\le{}s\le{}\min\{q,S\}\}$
\footnote{To avoid underflow and overflow, we also need to satisfy $S\ge{}A$, which is straightforwardly obtained by the existence of the feasible region $S(q)$ with $q$ setting as $Q$.}.

To transmit $s[n]$ packets in timeslot $n$, we determine the corresponding power consumption for the adaptive transmitter with the available Channel State Information (CSI).
In particular, we present the channel state $h[n]$ of timeslot $n$ by using the current channel coefficient of the fading channel.
As a result, we have that $h[n]$ belongs to the field of complex numbers $\mathbb{C}$.
With the channel state $h[n]$ given as $h\in\mathbb{C}$, we express the power consumption by function $P_{h}(s)$ for each transmission rate $s$, where we define function $P_{h}(s)=0$ for each $h$.
For typical communications scenarios, we provide a greater transmission rate by a greater power consumption, Meanwhile, the power efficiency will degrade with the increasing transmission rate [8].
Therefore, we focus on a function $P_h(s)$ that is monotonically increasing and convex in $s$ for each given $h$.
With $P_h(s)$ given for channel state $h[n]$, the power consumption in timeslot $n$
is defined as $\rho[n]=P_{h[n]}(s[n])$.

We further adopt an $L$-state block fading channel model, through which the channel coefficient of the fading channel stays invariant during each timeslot and is quantized into $L$ states, i.e., $h_1,h_2,\cdots,h_L$.
In this way, we have that channel state $h[n]$ belongs to set $\{h_1,\cdots,h_L\}$, through which we shall only consider the power functions $P_{h_\iota}(s),~\iota=1,\cdots,L$ for the block fading channel.
More specifically, the channel states are satisfy $0<|h_1|<|h_2|<\cdots<|h_L|<+\infty$.
In other words, we will obtain a better channel condition under a channel state $h_\iota$ with a greater index $l,~1\le{}\iota\le{}L$.
Moreover, we consider that the channel state $h[n]$ in each timeslot $n$ follows an independent and identically distributed (\emph{i.i.d.}) process.
As a result, we defined the probability of that channel state $h[n]$ for each timeslot $n$ is equal to $h_l$ as
\vspace{-2mm}
\begin{equation}
	\Pr\{h[n]=h_\iota\}=\eta_\iota,
\vspace{-4mm}
\end{equation}
where we have $\sum_{\iota=1}^L\eta_\iota=1$.

Under the cross-layer adaptive transmission policy, the transmission rate $s[n]$ is determined by the current queue length $q[n]$, the arrival rate $a[n]$, as well as the channel state $h[n]$.
With $q[n]$, $s[n]$, and $h[n]$ presented as $q$, $a$, $h_\iota$, respectively, we define the probability $f_{q,a,\iota}^s$ that transmission rate $s[n]$ is equal to $s$ as 
{\setlength\abovedisplayskip {2pt plus 7pt minus 8pt}
	\setlength\belowdisplayskip {2pt plus 7pt minus 8pt}
	\vspace{-2mm}
\begin{equation}\label{f_qa}
f_{q,a,\iota}^{s}=\Pr\{s[n]=s~|~q[n]=q,a[n]=a,h[n]=h_\iota\},
	\vspace{-2mm}
\end{equation}}%
where we have $\sum_{s=0}^{S}f_{q,a,\iota}^{s}=1$, and $f_{q,a,\iota}^s=0$ for each $s\notin\mathcal{S}(q)$.
Based on the probability $f_{q,a}^s$, the cross-layer adaptive transmission policy $\boldsymbol{F}$ is expressed by $\{f_{q,a,\iota}^{s}:0\le{}q\le{}Q,0\le{}a\le{}A,0\le{}\iota\le{}L,0\le{}s\le{}S\}$.
We first present the deterministic transmission policies using a degenerate probability distribution on the transmission rate for each given queue length, arrival rate and channel state.
Then, a deterministic policy $\boldsymbol{F}_D$ is equivalently expressed as $\{s^{\boldsymbol{F}_D}(q,a,\iota):0\le{}q\le{}Q,0\le{}a\le{}A,0\le{}\iota\le{}L\}$, where we have $s^{\boldsymbol{F}_D}(q,a,\iota)=\sum_{s\in\mathcal{S}(q)}sf_{q,a,\iota}^s$.
The set of deterministic policies are given by $\mathcal{F}_D\subsetneqq\mathcal{F}$, where $\mathcal{F}$ is the set of all policies.
For random arrivals that are temporally correlated, the same probabilistic strategy is also constructed by determining the probabilities of the transmission rate given the current queue length and channel state with the historical information of the arrival rates, as presented in Eq. (\ref{f_qa}).

By using the probabilistic transmission policies, we present a Markov Decision Process (MDP), where we express the system state as the triple $(q[n],a[n],h[n])$.
With system state $(q[n],a[n],h[n])$ at timeslot $n$ given as $(q,a,h_\iota)$, each adaptive transmission policy $\boldsymbol{F}\in\mathcal{F}$ can determine transmission rate $s[n]$ based on the probability distribution $\{f_{q,a,\iota}^s:0\le{}s\le{}S\}$.
Under the given transmission rate $s[n]$, we next determine the system state $(q[n+1],a[n+1],h[n+1])$ in timeslot $(n+1)$ following the processes of Markov arrival and channel fading.
In particular, the transition probability for the next timeslot is represented as
{\setlength\abovedisplayskip {2pt plus 7pt minus 8pt}
	\setlength\belowdisplayskip {1pt plus 7pt minus 8pt}
	\vspace{-2mm}
	\setlength\jot{0pt}
\begin{align}
\Pr\{q[n+1]=q',&a[n+1]=a',h[n+1]=h_{\iota'}~|\\\notag
&~q[n]=q,a[n]=a,h[n]=h_\iota,s[n]=s\}
=\gamma_{a,a'}\eta_{\iota'}\mathbbm{1}_{\{s=q+a'-q'\}},
\vspace{-4mm}
\end{align}}%
where we have $q'\in\{0,1,\cdots,Q\}$, $a'\in\{0,1,\cdots,A\}$, and $\iota'\in\{0,1,\cdots,L\}$.
With the system state employed, the MDP can continually evolve under the given initial queue length $q_0$, arrival rate $a_0$, and channel state $h_{\iota_0}$, where we define $q_0=q[0]$, $a_0=a[0]$, and $h_{\iota_0}=h[0]$.

With the formulated MDP, the long-term average power consumption and delay are also formulated based on the power consumption $\rho[n]=P_{h[n]}(s[n])$ and the queue length $q[n]$ in each timeslot, respectively.
First, the average power consumption $P_{\boldsymbol{F}}$ can be presented as
{
\vspace{-1mm}
\begin{equation}\label{ave_P}
	P_{\boldsymbol{F}}=\lim_{N\rightarrow\infty}\frac{1}{N}\mathbb{E}_{q_0,a_0,h_{\iota_0}}^{\boldsymbol{F}}\left\{\sum_{n=1}^{N}\rho[n]\right\},
	\vspace{-1mm}
\end{equation}}%
where $\mathbb{E}_{q_0,a_0,h_{\iota_0}}^{\boldsymbol{F}}\{\cdot\}$ is the expectation with respect to policy $\boldsymbol{F}$ as well as the initial system state $(q_0,a_0,h_{\iota_0})$.
The average delay $D_{\boldsymbol{F}}$ is given from Little's Law as
\vspace{-1mm}
\begin{equation}\label{ave_D}
D_{\boldsymbol{F}}=\lim_{N\rightarrow\infty}\frac{1}{N}\mathbb{E}_{q_0,a_0,h_{\iota_0}}^{\boldsymbol{F}}\left\{\frac{1}{\alpha}\sum_{n=1}^{N}q[n]\right\},
\vspace{-1mm}
\end{equation}%
where recall that $\alpha$ is defined as the expected number of packets that arrive in each timeslot.

Based on the average power consumption and delay in Eqs. (\ref{ave_P}) and (\ref{ave_D}), we can formulate the optimal delay-power tradeoff under Markov random arrivals.
Intuitively, a higher transmission rate can reduce the packets' delay, but degrades the power efficiency because $P_h(s)$ is convex on $s$ for each channel state $h$.
For a lower transmission rate, the reverse holds true, i.e., we have a greater power efficiency but also a larger transmission delay.
Therefore, a tradeoff exists between the delay and power consumption.
To obtain the optimal tradeoff, we formulate a cross-layer optimization problem as a Constrained Markov Decision Process (CMDP) under the probabilistic transmission strategy.
In the CMDP, we aim at minimizing the average delay subject to the constraint on the average power.
In particular, the optimization problem is \mbox{given as}%
{\setlength\abovedisplayskip {1pt plus 7pt minus 8pt}
\setlength\belowdisplayskip {1pt plus 7pt minus 8pt}
\setlength\jot{-1pt}
\begin{subequations}
\label{eqn_stationary_optimization}
\begin{align}
\min\limits_{\boldsymbol{F}\in\mathcal{F}}\quad & D_{\boldsymbol{F}}\\
\text{s.t.}\quad & P_{\boldsymbol{F}} \le P_{\text{th}}.
\end{align}%
\end{subequations}}%
By solving this CMDP under different power constraint $P_{\text{th}}$, we can show the optimal delay-power tradeoff under Markov arrivals.
As a result, we obtain the minimized average delay $D_{\boldsymbol{F}^\ast}$ and optimal policy $\boldsymbol{F}^\ast$ for each given $P_{\text{th}}$.

To particularly show the impact of Markovian arrivals, we first focus on the optimal delay-power tradeoff for an AWGN channel in Sections III and IV.
Then, we extend the optimal tradeoff by considering the fading channel in Section V.
More specifically, we present the AWGN channel by setting $L=1$ and $|h_1|=1$.
Under the only channel state, we further simplify the presentations of the only power function and the adaptive transmission policy as $P(s)$ and $\boldsymbol{F}=\{f_{q,a}^s:0\le{}q\le{}Q,0\le{}a\le{}A,0\le{}s\le{}S\}$ in the following two sections, respectively.
As a result, a degenerated CMDP is formulated with the system state as $(q[n],a[n])$.

\vspace{-5mm}
\section{{Optimal Delay-Power Tradeoff for AWGN Channels}}

\vspace{-2mm}

In this section, we focus on the optimal delay-power tradeoff for AWGN channels, which is described by the cross-layer optimization problem (\ref{eqn_stationary_optimization}).
We first show that the optimal delay-power tradeoff can be formulated by an equivalent LP problem based on the steady-state analysis for a single user.
With the LP problem being solved over the set of all the obtainable \mbox{power-delay} pairs, we then generate an optimal delay-power tradeoff curve for AWGN channels, under which minimized average delays are obtained for different power constraints.
Further, we show some interesting geometric properties of the optimal tradeoff curve.
Based on these geometric properties, we finally demonstrate that the same optimal tradeoff is obtained by the optimal policies with an arbitrary initial system state.
In other words, the optimal policies over AWGN channels have the same average delay and power consumptions regardless of the initial \mbox{system states.}

\vspace{-7mm}
\subsection{The equivalent LP problem}
\vspace{-3mm}

First, we show the optimal delay-power tradeoff by expressing the cross-layer optimization problem (\ref{eqn_stationary_optimization}) as an LP problem.
In particular, we formulate the LP problem based on a Markov Reward Process (MRP) that is generated by the CMDP with the transmission policy given.
For a given policy $\boldsymbol{F}$, we first describe the resulting MRP to analytically present the average delay and power.
In the MRP, $\lambda_{(q,a),(q',a')}$ denotes the transition probability from $(q,a)$ to $(q',a')$.
Based on the evolution of $q[n]$ and $a[n]$ in Eq. (\ref{queue_update}), transition probability $\lambda_{(q,a),(q',a')}$ is presented as
{\setlength\abovedisplayskip {2pt plus 7pt minus 8pt}
	\setlength\belowdisplayskip {2pt plus 7pt minus 8pt}
	\vspace{-2mm}
\begin{equation}
	\lambda_{(q,a),(q',a')}=
		\gamma_{a,a'}f_{q,a}^{q-q'+a'} \mathbbm{1}_{\{\max\{q-S,0\}\le{}q'-a'\le{}\min\{q,Q-A\}\}}.
\vspace{-2mm}
\end{equation}}%
With probability $\lambda_{(q,a),(q',a')}$, we then show steady-state probabilities by formulating \mbox{the balance} equations.
Let $\pi_{\boldsymbol{F}}(q,a)$ denote the steady-state probability. We present the balance equations as
{\setlength\abovedisplayskip {2pt plus 7pt minus 8pt}
	\setlength\belowdisplayskip {2pt plus 7pt minus 8pt}
	\vspace{-5mm}
\begin{equation}	\label{eq_balance}
	\sum_{a=0}^{A}\sum_{q=\max\{q'-a',0\}}^{\min\{q'-a'+S,Q\}}\pi_{\boldsymbol{F}}(q,a)\lambda_{(q,a),(q',a')}=\pi_{\boldsymbol{F}}(q',a'),
	\vspace{-3mm}
\end{equation}}%
where we have $\sum_{q=0}^{Q}\sum_{a=0}^{A}\pi_{\boldsymbol{F}}(q,a)=1$.
More specifically, $\pi_{\boldsymbol{F}}(q,a)$ indicates how often the queue length is equal to $q$ and the arrival rate is $a$ on average in the long run.
Considering the evolution of $q[n]$ in Eq. (\ref{queue_update}) with $s[n]\in\mathcal{S}(q[n])$, we have $q[n+1]-a[n+1]=q[n]-s[n]\le{}Q-A$ for each timeslot.
Therefore, it is straightforward that steady-state probability $\pi_{\boldsymbol{F}}(q,a)$ is equal to $0$ if $q-a>Q-A$.
By solving the balance equations for all queue lengths $q$ and arrival rates $a$, we can obtain the steady-state probability distribution $\boldsymbol{\pi}_{\boldsymbol{F}}$ that is defined as $\{\pi_{\boldsymbol{F}}(q,a):\forall{}q,a\}$.

The balance equations given by Eq. (\ref{eq_balance}) can be expressed as the following matrix form
{\setlength\abovedisplayskip {1pt plus 7pt minus 8pt}
	\setlength\belowdisplayskip {1pt plus 7pt minus 8pt}
	\vspace{-2mm}
	\begin{equation}
	\label{mat_balance_eq}
		\boldsymbol{\Lambda}_{\boldsymbol{F}}\boldsymbol{\pi}_{\boldsymbol{F}} = \boldsymbol{\pi}_{\boldsymbol{F}},
	\vspace{-2mm}
\end{equation}}%
where $\boldsymbol{\pi}_{\boldsymbol{F}}$ is formulated as vector with probabilities $\pi_{\boldsymbol{F}}(q,a)$ as elements.
In particular, we can present $\pi_{\boldsymbol{F}}(q,a)$ as the $\left(a\times(Q+1)+q+1\right)$th element in vector $\boldsymbol{\pi}_{\boldsymbol{F}}$.
Based on the permutation of $\pi_{\boldsymbol{F}}(q,a)$ in vector $\boldsymbol{\pi}_{\boldsymbol{F}}$, the stochastic matrix $\boldsymbol{\Lambda}_{\boldsymbol{F}}$ is also defined with $\lambda_{(q,a),(q',a')}$ as the elements.
The location of $\lambda_{(q,a),(q',a')}$ in $\boldsymbol{\Lambda}_{\boldsymbol{F}}$ is determined by the permutation of $\pi_{\boldsymbol{F}}(q,a)$ and $\pi_{\boldsymbol{F}}(q',a')$ in vector $\boldsymbol{\pi}_{\boldsymbol{F}}$.
In other words, when $\pi_{\boldsymbol{F}}(q,a)$ and $\pi_{\boldsymbol{F}}(q',a')$ are the $i$th and $j$th elements in $\boldsymbol{\pi}_{\boldsymbol{F}}$, respectively, we have $\lambda_{(q,a),(q',a')}$ is located at the $i$th column and $j$th row in matrix $\boldsymbol{\Lambda}_{\boldsymbol{F}}$.

By using the steady-state probability, we next present the average power consumption and delay.
Given the steady-state probability $\pi_{\boldsymbol{F}}(q,a)$, we express the average power \mbox{consumption as}%
{\setlength\abovedisplayskip {2pt plus 7pt minus 8pt}
	\setlength\belowdisplayskip {2pt plus 7pt minus 8pt}
	\vspace{-4.5mm}
\begin{equation}
P_{\boldsymbol{F}}
=\sum_{q=0}^{Q}\sum_{a=0}^{A}\sum_{s=0}^{S}P(s)\pi_{\boldsymbol{F}}(q,a)f_{q,a}^s.
\label{PwithPi}
\vspace{-1mm}
\end{equation}}%
Similarly, the average delay is given as
{\setlength\abovedisplayskip {2pt plus 7pt minus 8pt}
	\setlength\belowdisplayskip {2pt plus 7pt minus 8pt}
	\vspace{-1mm}
\begin{equation}
D_{\boldsymbol{F}}=\frac{1}{\alpha}\sum_{q=0}^{Q}\sum_{a=0}^{A} q \pi_{\boldsymbol{F}}(q,a).
\label{DwithPi}
	\vspace{-1mm}
\end{equation}}%


Then, we demonstrate the optimal delay-power tradeoff under the cross-layer transmission policies.
As shown in Eqs. (\ref{PwithPi}) and (\ref{DwithPi}), the average power consumption and delay are presented based on the steady-state probability $\pi_{\boldsymbol{F}}(q,a)$ with policy $\boldsymbol{F}$ given.
Considering the steady-state probabilities that satisfy the balance equations in Eq. (\ref{mat_balance_eq}), we can reveal the optimal delay-power tradeoff given by problem (\ref{eqn_stationary_optimization}) by the solution in the following problem for each value of $P_{\rm{th}}$.%
{\setlength\abovedisplayskip {2pt plus 7pt minus 8pt}
	\setlength\belowdisplayskip {2pt plus 7pt minus 8pt}	
\setlength\jot{-1.5pt}
\begin{subequations}
\label{eqn_cmdp_expand}
\begin{align}
\min\limits_{\{\boldsymbol{\pi}_{\boldsymbol{F}},\boldsymbol{F}\}}\quad
& \frac{1}{\alpha}\sum_{q=0}^{Q}\sum_{a=0}^{A} q \pi_{\boldsymbol{F}}(q,a)\label{eqn_cmdp_expand_0}\\
\text{s.t.}\quad
& \sum_{q=0}^{Q}\sum_{a=0}^{A}\sum_{s=0}^{S}P(s)\pi_{\boldsymbol{F}}(q,a)f_{q,a}^s\le{}P_{\textrm{th}} \label{eqn_cmdp_expand_1}\\
& \boldsymbol{\Lambda}_{\boldsymbol{F}}\boldsymbol{\pi}_{\boldsymbol{F}} = \boldsymbol{\pi}_{\boldsymbol{F}} \label{eqn_cmdp_expand_2}\\
& \sum_{q=0}^{Q}\sum_{a=0}^{A}\pi_{\boldsymbol{F}}(q,a)=1 \label{eqn_cmdp_expand_3}\\
& \sum_{s=0}^S f_{q,a}^s=1 \qquad \forall~q,~a \label{eqn_cmdp_expand_4}\\
&\pi_{\boldsymbol{F}}(q,a)\ge 0,~ f_{q,a}^{s}\ge 0\qquad \forall~q,~a,~s,\label{eqn_cmdp_expand_5}
\end{align}
\end{subequations}}%
where the optimal delay for the problem is generated by the optimal transmission policy $\boldsymbol{F}^\ast$ with the corresponding steady-state probability $\pi^\ast_{\boldsymbol{F}^\ast}(q,a)$.

With the cross-layer optimization problem (\ref{eqn_cmdp_expand}) given, we finally convert problem (\ref{eqn_cmdp_expand}) to an equivalent LP problem, through which the optimal average delay is obtained for each given power constraint $P_{\rm{th}}$.
To formulate the LP problem, we use the product of $\pi_{\boldsymbol{F}}(q,a)$ and $f_{q,a}^{s}$ as the optimization variables.
Defining $x_{q,a}^s$ as $\pi_{\boldsymbol{F}}(q,a)f_{q,a}^{s}$, we can present the optimal delay-power tradeoff by using that equivalent LP problem that is shown in the following theorem.
\vspace{-3mm}
\begin{theorem}\label{th_cmdp_lp}
	The problem (\ref{eqn_cmdp_expand}) is equivalent to the following linear programming problem.%
	{\setlength\abovedisplayskip {1pt plus 7pt minus 8pt}
		\setlength\belowdisplayskip {1pt plus 7pt minus 8pt}
	\setlength\jot{-1.5pt}
	\begin{subequations}
		\label{eqn_lp}
		\begin{align}
		\min\limits_{\{x_{q,a}^{s}\}} \quad
		& \frac{1}{\alpha}\sum_{q=0}^{Q} \sum_{a=0}^{A} \sum_{s=0}^S q x_{q,a}^{s} \label{obj_1}\\
		\text{s.t.}\quad
		& \sum_{q=0}^{Q} \sum_{a=0}^{A} \sum_{s=0}^S P(s) x_{q,a}^{s} \le P_{\text{th}} \label{con_P}\\
		&\sum_{q=\max\{q'-a',0\}}^{\min\{q'-a'+S,Q\}}\sum_{a=0}^{A}\sum_{s=0}^{S}\gamma_{a,a'}x_{q,a}^{s}\mathbbm{1}_{\{s=q+a'-q'\}}=\sum_{s=0}^{S}x_{q',a'}^{s}\nonumber\\
		&\forall~0\le{}q'\le{}Q,~0\le{}a'\le{}A \label{con_1}\\
		& \sum_{q=0}^{Q} \sum_{a=0}^{A} \sum_{s=0}^S x_{q,a}^{s}=1 \label{con_2}\\
		& x_{q,a}^s\ge 0 \qquad \forall~0\le{}q\le{}Q,~0\le{}a\le{}A,~0\le{}s\le{}S \label{con_3}.
		\end{align}
	\end{subequations}}%
\end{theorem}
\vspace{-5mm}
\begin{IEEEproof}
	To show the equivalence of problems (\ref{eqn_cmdp_expand}) and (\ref{eqn_lp}), we divide the proof into two parts.
	We first show that problem (\ref{eqn_cmdp_expand}) is converted into LP problem (\ref{eqn_lp}) by replacing $\pi_{\boldsymbol{F}}(q,a)f_{q,a}^{s}$ as $x_{q,a}^{s}$.
	For each feasible solution $\boldsymbol{\pi}_{\boldsymbol{F}}$ and $\boldsymbol{F}$ of problem (\ref{eqn_cmdp_expand}), we can generate a feasible solution for problem (\ref{eqn_lp}), i.e., $\{x_{q,a}^s=\pi_{\boldsymbol{F}}(q,a)f_{q,a}^{s}\}$.
	By using the corresponding $\{x_{q,a}^s\}$ in problem (\ref{eqn_lp}), we also obtain the same average power consumption and delay as $\boldsymbol{\pi}_{\boldsymbol{F}}$ and $\boldsymbol{F}$ in problem (\ref{eqn_cmdp_expand}).
	
	For each feasible solution $\{x_{q,a}^s\}$ of problem (\ref{eqn_lp}), we then construct the corresponding policy $\boldsymbol{F}$ by presenting probability $f_{q,a}^{s}$ as %
		{\setlength\abovedisplayskip {2pt plus 6pt minus 8pt}
			\setlength\belowdisplayskip {2pt plus 6pt minus 8pt}
			\vspace{-2mm}
		\begin{equation}
		f_{q,a}^{s}=\left\{
		\begin{array}{ll}
		\frac{x_{q,a}^{s}}{\pi_{\boldsymbol{F}}(q,a)}, & \pi_{\boldsymbol{F}}(q,a)>0,\\
		\mathds{1}_{\{s=\min\{q,S\}\}}, & \pi_{\boldsymbol{F}}(q,a)=0,
		\end{array}\right.
		\label{x_to_F}
		\vspace{-1mm}
		\end{equation}}%
	where steady-state probability $\pi_{\boldsymbol{F}}(q,a)$ under policy $\boldsymbol{F}$ is expressed as $
		\pi_{\boldsymbol{F}}(q,a)=\sum_{s=0}^{S}x_{q,a}^{s}.$
	By substituting the attained $f_{q,a}^s$ and $\pi_{\boldsymbol{F}}(q,a)$ into problem (\ref{eqn_cmdp_expand}), we can check that the constructed solution satisfies the balance equations in Eq. (\ref{eqn_cmdp_expand_2}) with the average delay and power consumption remain unchanged, through which we complete the proof.%
\end{IEEEproof}

With the equivalent LP problem (\ref{eqn_lp}) formulated, we demonstrate the optimal delay-power tradeoff for AWGN channels.
By solving the derived LP problem, we can particularly obtain the minimum average delay with the optimal policy $\boldsymbol{F}^\ast$ given by Eq. (\ref{x_to_F}).

\vspace{-6mm}
\subsection{The Optimal Delay-Power Tradeoff Curve}
\vspace{-3mm}

In this subsection, we attain the optimal delay-power tradeoff channel by solving the LP problem (\ref{eqn_lp}) that is formulated in Theorem \ref{th_cmdp_lp} for AWGN channels.
By solving the LP problem over a power-delay plane that contains all the obtainable power and delay pairs under the policies, we present the optimal delay-power tradeoff curve.
In this way, the minimized average delay can be obtained for the single link under a given average power constraint.

To obtain the optimal delay-power tradeoff curve, we first solve LP problem (\ref{eqn_lp}) by considering the set of all obtainable average power-delay pairs.
In particular, a power-delay plane is first formulated to contain all the average power-delay pairs $(P_{\boldsymbol{F}},D_{\boldsymbol{F}})$ that are generated by the cross-layer transmission policies $\boldsymbol{F}\in\mathcal{F}$.
However, for a given transmission policy $\boldsymbol{F}=\{f_{q,a}^s\}$, we can only present the corresponding average power-delay pair $(P_{\boldsymbol{F}},D_{\boldsymbol{F}})$ by $f_{q,a}^s$ with the assistant of $\pi_{\boldsymbol{F}}(q,a)$ as Eqs. (\ref{PwithPi}) and (\ref{DwithPi}), respectively.
Considering we determine $\pi_{\boldsymbol{F}}(q,a)$ under policy $\boldsymbol{F}$ based on the a series of balance equations in Eq. (\ref{eq_balance}), we can hardly show the power-delay pair $(P_{\boldsymbol{F}},D_{\boldsymbol{F}})$ as a analytical expression of $f_{q,a}^s$.
In this way, we generate the power-delay plane based on the optimization variables $x_{q,a}^s$ in LP problem (\ref{eqn_lp}), which can be referred to as the state-action frequency in MDP \cite[Section 8.9]{puterman2014markov}.
With the obtainable state-action frequencies $\{x_{q,a}^s\}$ given, we can analytically present the average power-delay pair by the objective function and power constraint in LP problem (\ref{eqn_lp}).
The corresponding policy $\boldsymbol{F}$ is also obtained following the bijective map presented in Theorem \ref{th_cmdp_lp}.

Thus, we first express the set that consists of all the obtainable state-action frequencies $\{x_{q,a}^{s}\}$
under the transmission policies as
{\setlength\abovedisplayskip {6pt plus 6pt minus 8pt}
	\setlength\belowdisplayskip {6pt plus 6pt minus 8pt}
	\vspace{-4mm}
\begin{equation}
	\mathcal{G}=\left\{
	\{x_{q,a}^{s}:\forall~q,~a,~s\}~|~
\text{Eqs. (\ref{con_1}), (\ref{con_2}), and (\ref{con_3}})
	\right\}.
	\vspace{-4mm}
\end{equation}}%
According to the linear functions in objective function (\ref{obj_1}) and power constraint (\ref{con_P}), we then present the average delay and power, respectively, for the feasible $\{x_{q,a}^s\}$.
As a result, the power-delay plane is generated to contain all the obtainable average power-delay pairs.

We then express the feasible state-action frequencies $\{x_{q,a}^{s}\}$ as a $((Q+1)\times(A+1)\times(S+1))$-dimension vector.
We can straightforwardly demonstrate set $\mathcal{G}$ as a polyhedron in a high dimensional Euclidean space.
The obtainable power-delay pairs are next presented as the projection of the state-action frequencies on the power-delay plane.
In other words, the set $\mathcal{R}$ of all the obtainable average power-delay pairs \mbox{is defined as}
{\setlength\abovedisplayskip {6pt plus 6pt minus 8pt}
	\setlength\belowdisplayskip {6pt plus 6pt minus 8pt}
	\vspace{-2mm}
\begin{equation}
\mathcal{R}=\left\{(P,D)~|~\forall \{x_{q,a}^s\}\in\mathcal{G},~P=\sum_{q=0}^{Q} \sum_{a=0}^{A} \sum_{s=0}^S P(s) x_{q,a}^{s},~D = \frac{1}{\alpha}\sum_{q=0}^{Q} \sum_{a=0}^{A} \sum_{s=0}^S q x_{q,a}^{s}\right\},
\label{set_R}
	\vspace{-2mm}
\end{equation}}%
where set $\mathcal{R}$ is a polyhedron on the power-delay plane.



With definition of set $\mathcal{R}$ in Eq. (\ref{set_R}), we rewrite the LP problem (\ref{eqn_lp}) over the power-delay plane.
In particular, we have
{\setlength\abovedisplayskip {1pt plus 2pt minus 3pt}
	\setlength\belowdisplayskip {1pt plus 2pt minus 3pt}
 \setlength\jot{-2pt}
 \vspace{-4mm}
\begin{subequations}
	\label{eqn_P_D_plane}
	\begin{align}
		\min\limits_{(P,D)\in\mathcal{R}}\quad
		& D\\
		\text{s.t.}\quad
		& P\le{}P_{\text{th}}.
	\end{align}
\end{subequations}}%
In this way, we demonstrate the optimal delay-power tradeoff described in cross-layer optimization problem (\ref{eqn_stationary_optimization}) over the power-delay plane.
With the derived LP problem in Eq. (\ref{eqn_P_D_plane}), we obtain the optimal power-delay pair $(P^\ast,D^\ast)$ by searching the power-delay pair that minimizes the delay in set $\mathcal{R}\cap\{(P,D)~|~P\le{}P_{\text{th}}\}$.

We finally formulate the optimal delay-power tradeoff curve for AWGN channels as
{\setlength\abovedisplayskip {1pt plus 2pt minus 3pt}
	\setlength\belowdisplayskip {1pt plus 2pt minus 3pt}
	 \vspace{-2mm}
\begin{equation}\label{curve_D_P}
\mathcal{L}=\{(P^\ast,D^\ast)\in\mathcal{R}~|~\forall(\bar{P},\bar{D})\in\mathcal{R},\text{ either }P^\ast\le{}\bar{P}\text{ or }D^\ast\le{}\bar{D}\},
 \vspace{-2mm}
\end{equation}}%
which consists of all the optimal delay-power pairs under different power constraints.
For each optimal power-delay pair $(P^\ast,D^\ast)$ in problem (\ref{eqn_P_D_plane}), $(P^\ast,D^\ast)$ belongs to $\mathcal{L}$ because we have that $D^\ast\le\bar{D}$ if $(\bar{P},\bar{D})\in\mathcal{R}\cap\{\bar{P}\le{}P_{\text{th}}\}$, and $P^\ast\le{}P_{\text{th}}\le\bar{P}$ if $(\bar{P},\bar{D})\in\mathcal{R}\cap\{\bar{P}\ge{}P_{\text{th}}\}$.
Meanwhile, each element $(P^\ast,D^\ast)$ in set $\mathcal{L}$ can minimize the average delay in problem (\ref{eqn_P_D_plane}) with power constraint $P_{\text{th}}$ as $P^\ast$.
Further, the geometric properties of the optimal delay-power tradeoff curve are then presented in the following theorem.
\vspace{-3mm}
\begin{theorem}
The optimal tradeoff curve $\mathcal{L}$ is piecewise linear, decreasing, and convex.
\label{theorem_piecewise_linear}
\end{theorem} 
\vspace{-4mm}
\begin{IEEEproof}
	The proof of the geometric properties follows directly from \cite[Corollary 3]{chen2017delaytcom}.
	We include the main idea of it for completeness.
	With the optimal tradeoff curve $\mathcal{L}$ expressed as Eq. (\ref{curve_D_P}), we first show that $\mathcal{L}$ is convex and decreasing according to the definitions of convex and decreasing function, respectively.
	By showing $\mathcal{L}$ as a part of bound of the polyhedron $\mathcal{R}$, we next present $\mathcal{L}$ as a piecewise linear curve.
	\end{IEEEproof}
\vspace{-1mm}

In this way, we present the optimal delay-power tradeoff by solving the equivalent LP problem on the power-delay plane.
By employing the state-action frequencies, we analytically present the optimal delay-power tradeoff curve for AWGN channels, under which the minimized average delay is attained for the adaptive transmitter with a given power constraint.

\vspace{-6mm}
\subsection{The Optimal Delay-Power Tradeoff with an Arbitrary Initial State}
\vspace{-2mm}

In this subsection, we show that the same optimal delay-power tradeoff is obtained for AWGN channels by the optimal adaptive transmission policies under an arbitrary initial state.
With different initial queue lengths and arrival rates, we may have different average delays and powers under a given transmission policy because different steady-state distributions can be obtained with multiple closed classes existing in the corresponding MRP \cite[Section 4.3]{kao1997introduction}.
However, for the optimal transmission policies of LP problem (\ref{eqn_lp}), we show that the same average delay and power consumption is obtained for AWGN channels with an arbitrary initial state.

For each power-delay pair on curve $\mathcal{L}$, the corresponding optimal adaptive transmission policy is first formulated by solving LP problem (\ref{eqn_lp}).
In particular, with the optimal solution $\{{x^\ast}_{q,a}^s\}$ of LP problem (\ref{eqn_lp}), we obtain the optimal policy by determining $f_{q,a}^s$ according to Eq. (\ref{x_to_F}).

Then, we demonstrate that the optimal adaptive transmission policy can obtain the same optimal tradeoff under an arbitrary initial state.
In other words, we show that the performance of the optimal policy on the average delay and power consumption is independent with an initial state.
For this purpose, we only need to show that the Markov chain induced by the MRP has only one closed communication class under an optimal policy.
These Markov chains are referred to as unichain.
First, we present the structure of the Markov chains for the vertices of the optimal delay-power tradeoff curve $\mathcal{L}$ in the following theorem.%
\vspace{-4mm}
\begin{theorem}
	The optimal delay-power tradeoff curve $\mathcal{L}$ satisfies that
	\vspace{-1mm}
	\begin{enumerate}
		\item All vertices of $\mathcal{L}$ can be obtained by adaptive transmission policies with unichains;
		\item All vertices of $\mathcal{L}$ can be obtained by deterministic transmission policies;
		\item The policies corresponding to two adjacent vertices of $\mathcal{L}$ have different transmission rates only on one state.
	\end{enumerate}
	\label{theorem_tradeoff_curve_policy_vertex}
\end{theorem}%
\vspace{-5mm}
\begin{IEEEproof}
	See Appendix \ref{Appendix_A}.
\end{IEEEproof}
\vspace{-1mm}

The vertices of the optimal tradeoff curve $\mathcal{L}$ can be generated by the optimal deterministic transmission policies, under which the Markov chains have only one closed class.
As a result, for all the vertices of curve $\mathcal{L}$, the same optimal delay-power tradeoff can be presented by the corresponding optimal transmission policies for any arbitrary initial state.

We next show that the same minimized average delay can be obtained under an arbitrary initial state for the other power-delay pairs on $\mathcal{L}$.
Since $\mathcal{L}$ is piecewise linear, we first consider the optimal power-delay points by dividing the curve into several segments with a pair of adjacent vertices as endpoints.
By using the two adaptive transmission policies for the pair of adjacent vertices, we then construct the optimal policies with unichains for each segment of curve $\mathcal{L}$.
In particular, the construction of the optimal policies relies on the following lemma.
\vspace{-4mm}
\begin{lemma}
	$\boldsymbol{F}'=\{{f'}_{q,a}^s\}$ and $\boldsymbol{F}''=\{{f''}_{q,a}^s\}$ are two transmission policies with unichains, and have different distributions on the transmission rate only when $q=\tilde{q}$ and $a=\tilde{a}$.
	We define policy $\boldsymbol{F}=\epsilon\boldsymbol{F}'+(1-\epsilon)\boldsymbol{F}''$, where each $f_{q,a}^s$ is equal to $\epsilon{f'}_{q,a}^s+(1-\epsilon){f''}_{q,a}^s$, and $0\le \epsilon\le 1$.
	Then, we have
	\vspace{-2mm}
	\begin{enumerate}
		\item The Markov chain under policy $\boldsymbol{F}\!=\!\epsilon\boldsymbol{F}'\!+\!(1-\epsilon)\boldsymbol{F}''$ is a unichain for each $0\le \epsilon\le 1$;
		\item There exists a $0\!\le \!\epsilon'\!\le \!1$ so that $P_{\boldsymbol{F}}\!=\!\epsilon'P_{\boldsymbol{F}'}\!+\!(1-\epsilon') P_{\boldsymbol{F}''}$ and $D_{\boldsymbol{F}}\!=\!\epsilon'D_{\boldsymbol{F}'}\!+\!(1-\epsilon' )D_{\boldsymbol{F}''}$;
		\item Parameter $\epsilon'$ increasingly moves from $0$ to $1$ with the increase of $\epsilon$ from interval $[0,1]$.
	\end{enumerate}%
	\label{lemma_linearcombination}%
\end{lemma}%
\vspace{-5mm}
\begin{IEEEproof}
	See Appendix \ref{Appendix_C}.
\end{IEEEproof}
\vspace{-1mm}

For each pair of adjacent vertices $(\hat{P},\hat{D})$ and $(\tilde{P},\tilde{D})$ on $\mathcal{L}$, we present the two optimal deterministic policies $\boldsymbol{\hat{F}}^\ast$ and $\boldsymbol{\tilde{F}}^\ast$ with unichains, according to \mbox{Theorem \ref{theorem_tradeoff_curve_policy_vertex}}.
The pair of policies has different transmission rates only for one particular queue length and arrival rate.
According to Lemma \ref{lemma_linearcombination}, we can present the optimal policy $\boldsymbol{F}^\ast$ as $(1-\epsilon)\boldsymbol{\hat{F}}^\ast+\epsilon\boldsymbol{\tilde{F}}^\ast$, by which the average power-delay is presented as $(\epsilon'\hat{P}\!+\!(1-\epsilon') \tilde{P},\epsilon'\hat{D}\!+\!(1-\epsilon' )\tilde{D})$, and the Markov chain is a unichain.
As a result, we show the existence of the optimal policy with a unichain for each power-delay pairs $(P,D)$ on the optimal delay-power tradeoff curve $\mathcal{L}$.

According to Theorem \ref{theorem_tradeoff_curve_policy_vertex} and Lemma \ref{lemma_linearcombination}, we finally straightforwardly show that the optimal delay-power tradeoff curve is obtained under an arbitrary initial state in the following theorem.
\vspace{-2mm}
\begin{theorem}
	All the average power-delay pairs of the optimal delay-power tradeoff curve $\mathcal{L}$ can be obtained using the adaptive transmission policies with unichains.
	\label{theorem_tradeoff_curve_policy}
\end{theorem}
\vspace{-2mm}

Therefore, the optimal delay-power tradeoff for AWGN channels is obtained by the optimal policy that is given by the LP problem (\ref{eqn_lp}).
Meanwhile, the same optimal tradeoff is presented for the single link with different initial queue lengths and arrival rates.

\vspace{-3mm}
\section{Threshold-based Optimal Transmission Policy over AWGN channels}
\vspace{-2mm}

In this section, we show the threshold-based structure for the optimal adaptive transmission policies over AWGN channels.
For each optimal average power-delay pair, we present the delay-optimal transmission strategy by using a threshold-based structure on the queue length, in which the thresholds for different transmission rates are given for the arrival rates.
To this end, we first present the threshold-based optimal policies for the vertices of the optimal tradeoff curve $\mathcal{L}$ based on the Lagrangian relaxation of the cross-layer optimization problem (\ref{eqn_stationary_optimization}).
Further, by using the optimal policies on the vertices, we formulate the threshold-based transmission policy for each average power-delay pairs on curve $\mathcal{L}$.
With the threshold-based structure, we finally develop a threshold-based algorithm to efficiently obtain the optimal delay-power tradeoff.

\vspace{-4mm}
\subsection{Threshold-based Optimal Deterministic Policy for the Lagrangian Relaxation Problem}
\vspace{-3mm}

\begin{figure}[t]
	\centering
	\includegraphics[width=0.50\columnwidth, height=0.41\columnwidth]{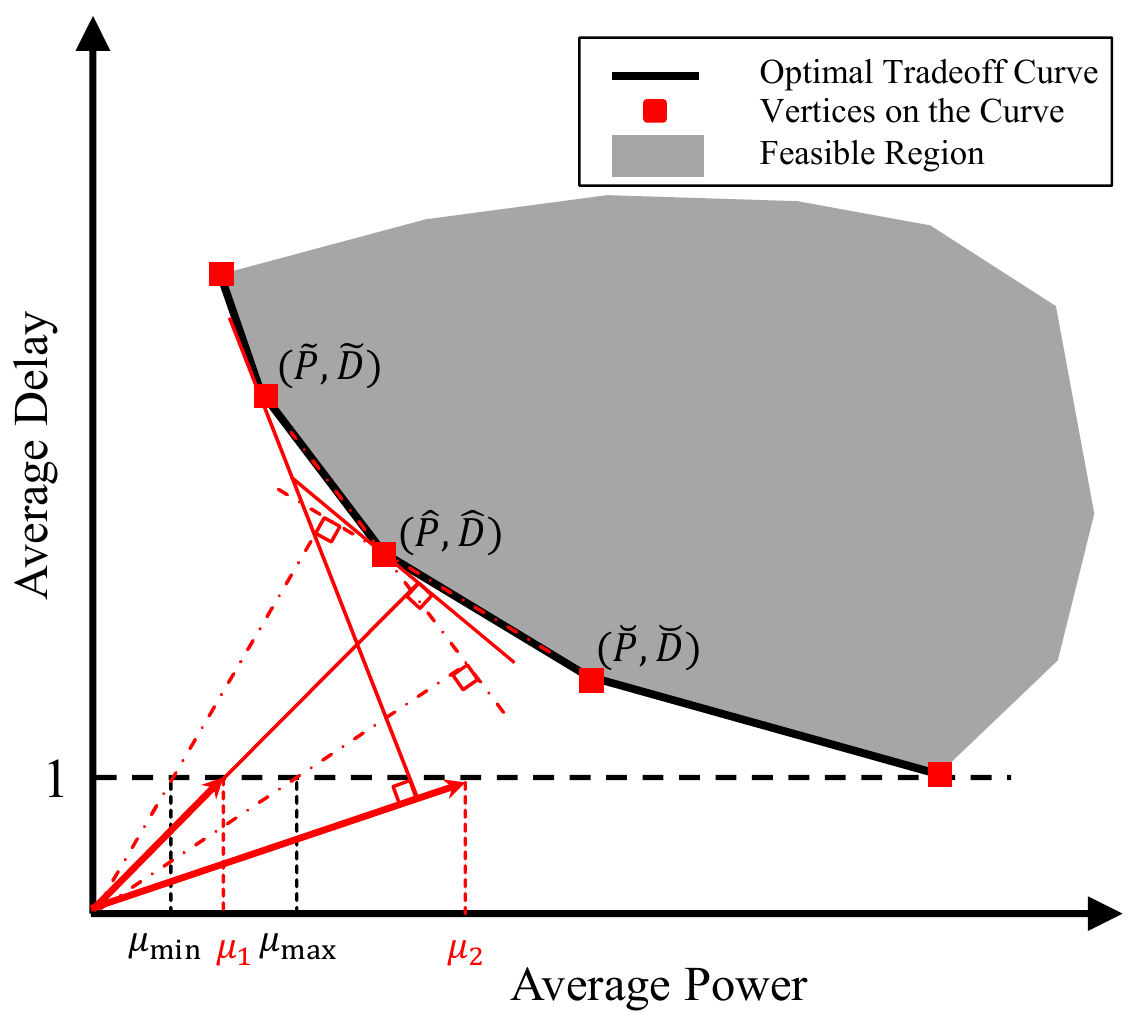}
	\vspace{-5mm}
	\caption{The stretch of the optimal delay-power tradeoff: We present the vertex of the curve as $(\hat{P},\hat{D})$, while two vertices that are adjacent with vertex $(\hat{P},\hat{D})$ are $(\tilde{P},\tilde{D})$ and $(\breve{P},\breve{D})$ with $\tilde{P}<\hat{P}<\breve{P}$.
			As for the two end points of the curve, we have no $(\tilde{P},\tilde{D})$ for the vertex with lowest power; no $(\breve{P},\breve{D})$ for the vertex with largest power.}
	\vspace{-4mm}
	\label{fig_weightedsum_piecewiselinear}
\end{figure}

In this subsection, the threshold-based optimal deterministic policies are shown for all the vertices of the optimal delay-power tradeoff curve $\mathcal{L}$ that is formulated for AWGN channels.
For each vertex on $\mathcal{L}$, we first obtain an optimal deterministic policy by exploiting the Lagrangian relaxation problem for cross-layer optimization problem (\ref{eqn_stationary_optimization}).
Then, for the optimal deterministic policies, we show that there exists a threshold-based structure on the queue lengths.

First, we formulate the Lagrangian relaxation problem for each vertex.
As shown in \figurename~\ref{fig_weightedsum_piecewiselinear}, for each $\mu>0$, we always find a vertex on tradeoff curve $\mathcal{L}$ to get the minimum value of $D+\mu{}P$.
For each vertex on $\mathcal{L}$, we further show a set of $\mu$ as $(\mu_{\min},\mu_{\max})$ \footnote{When $\mu$ is equal to $\mu_{\min}$ or $\mu_{\max}$, two adjacent vertices can obtain the minimum value of $D+\mu{}P$.}, under which the vertex obtains the minimized value of $D+\mu{}P$.
In particular, we have that $\mu_{\min}=\frac{\breve{P}-\hat{P}}{\hat{D}-\breve{D}}$ for all the vertices except the one with the largest power, while we set $\mu_{\min}$ as $0$ for this vertex based on the observation of Fig. \ref{fig_weightedsum_piecewiselinear}.
Similarly, we have $\mu_{\max}=\frac{\hat{P}-\tilde{P}}{\tilde{D}-\hat{D}}$ for the vertices with a less power, and $\mu_{\max}=+\infty$ for the vertex with the lowest power.

Since set $\mathcal{R}$ consists of all the power-delay pairs given by policies $\boldsymbol{F}\in\mathcal{F}$, we can show the optimal policy for each vertex by the following Lagrangian relaxation problem
\vspace{-1mm}
{\setlength\abovedisplayskip {2pt plus 5pt minus 6pt}
	\setlength\belowdisplayskip {2pt plus 5pt minus 6pt}
\begin{align}
\min\limits_{\boldsymbol{F}\in \mathcal{F}} \quad D_{\boldsymbol{F}}+\mu P_{\boldsymbol{F}}-\mu P_{\text{th}},
\label{eqn_lag_relax}
\end{align}}%
where the multiplier $\mu$ belongs to the corresponding set for the given vertex $(\hat{P},\hat{D})$ on $\mathcal{L}$.

Therefore, we show the optimal policy for each vertex $(\hat{P},\hat{D})$ by solving Lagrangian relaxation problem (\ref{eqn_lag_relax}) with specific $\mu$ employed.
In particular, we formulate problem (\ref{eqn_lag_relax}) as an unconstrained infinite-horizon MDP with the objective function as $D_{\boldsymbol{F}}+\mu{}P_{\boldsymbol{F}}$.
According to the result given by \cite[Theorem 9.1.8]{puterman2014markov}, we have that the unconstrained MDP is minimized by a deterministic policy, under which the corresponding Markov chain is a unichain.
Further, we show that the deterministic policy is presented by a threshold-based structure on the \mbox{queue length.}%
\vspace{-3mm}
\begin{theorem}
For each vertex $(\hat{P},\hat{D})$ on curve $\mathcal{L}$, the optimal deterministic policy $\boldsymbol{F}^\ast$ is presented by the threshold-based structure on the queue length, in which thresholds $q_{\boldsymbol{F}^\ast}(s,a)$ exist for every $0\le{}s\le{}S,~0\le{}a\le{}A$, and the probabilities ${f^\ast}_{q,a}^s$ satisfy that
{\setlength\abovedisplayskip {4pt plus 5pt minus 6pt}
	\setlength\belowdisplayskip {4pt plus 5pt minus 6pt}
	\setlength\jot{1pt}
	\begin{equation}
	\left\{
	\begin{array}{ll}
	{f^\ast}_{q,a}^s=1 & q_{\boldsymbol{F}^\ast}(s-1,a)<q\le q_{\boldsymbol{F}^\ast}(s,a), \\
	{f^\ast}_{q,a}^s=0 & \text{otherwise},
	\end{array}\right.
	\label{eqn_deterministic_threshold}
	\end{equation}}%
	where we have $0\!\le\!{}q_{\boldsymbol{F}^\ast}(0,a)\!\le\!{}q_{\boldsymbol{F}^\ast}(1,a)\!\le\!{}\cdots\!\le\!{}q_{\boldsymbol{F}^\ast}(S,a)\!\le\!{}Q$ and $q_{\boldsymbol{F}^\ast}(-1,a)\!=\!-1$ for each $a$.
	\label{theorem_01}
\end{theorem}
\vspace{-3mm}
\begin{IEEEproof}
	See Appendix \ref{Appendix_B}.
\end{IEEEproof}
\vspace{-1mm}

With a threshold-based optimal deterministic policy $\boldsymbol{F}^\ast$ given, we show a series of thresholds $\{q_{\boldsymbol{F}^\ast}(s,a):s=0,1,\cdots,S\}$ for each arrival rate $a\in\{0,1,\cdots,A\}$.
By using the thresholds on the queue length, we then can completely describe the corresponding optimal deterministic policy for each vertex of the optimal tradeoff curve $\mathcal{L}$.
Moreover, we can determine the delay-optimal transmission strategy by using the order relation of queue lengths with the thresholds under different arrival rates.

\vspace{-3mm}
\subsection{Threshold-Based Optimal Adaptive Transmission Policy}
\vspace{-1mm}

We now present the threshold-based optimal policy for each power-delay pair on the optimal delay-power tradeoff curve $\mathcal{L}$. 
With a given power-delay pair on curve $\mathcal{L}$, we construct the threshold-based optimal policy as a convex combinations of the optimal deterministic policies for the vertices on $\mathcal{L}$ which are presented in Theorem \ref{theorem_01}.
In particular, we present the threshold-based optimal policies for AWGN channels in the following theorem.%
\vspace{-3mm}
\begin{theorem}
The optimal policy $\boldsymbol{F}^\ast$ exists $(A+1)\times(S+1)$ thresholds $q_{\boldsymbol{F}^\ast}(s,a)$, where we have $0\le{}q_{\boldsymbol{F}^\ast}(0,a) \le q_{\boldsymbol{F}}(1,a) \le \cdots \le q_{\boldsymbol{F}}(S,a)\le{}Q$ for each arrival rate $a=0,1,\cdots,A$.
With all the thresholds $q_{\boldsymbol{F}^\ast}(s,a)$ given, the optimal policy $\boldsymbol{F}^\ast$ satisfies
{\setlength\abovedisplayskip {1pt plus 5pt minus 6pt}
	\setlength\belowdisplayskip {1pt plus 5pt minus 6pt}
	\setlength\jot{0pt}
	\vspace{-1mm}
	\begin{equation}\label{eqn_threshold}
		\left\{
		\begin{array}{ll}
			{f^\ast}_{q,a}^s=1 & q_{\boldsymbol{F}^\ast}(s-1,a)<q\le q_{\boldsymbol{F}^\ast}(s,a), a\neq a^\ast ~\text{or}~ s\neq s^\ast\\
			{f^\ast}_{q,a}^s=1 & q_{\boldsymbol{F}^\ast}(s-1,a)<q < q_{\boldsymbol{F}^\ast}(s,a),a= a^\ast ~\text{and}~ s= s^\ast\\
			{f^\ast}_{q,a}^s+{f^\ast}_{q,a}^{(s-1)}=1&q = q_{\boldsymbol{F}^\ast}(s,a),a= a^\ast~ \text{and}~ s= s^\ast\\
			{f^\ast}_{q,a}^s=0 & \text{otherwise}.
		\end{array}
		\right.
	\end{equation}
}%
	where the specific transmission rate $s^\ast$ and arrival rate $a^\ast$ are given by optimal policy $\boldsymbol{F}^\ast$, and we have $q_{\boldsymbol{F}^\ast}(-1,a)=-1$ for each $0\le{}a\le{}A$.
	\label{theorem_threshold}
\end{theorem}
\vspace{-4mm}
\begin{IEEEproof}
Our proof starts with the observation that the optimal policies corresponding to the vertices of the optimal delay-power tradeoff curve $\mathcal{L}$ satisfy Eq. (\ref{eqn_threshold}).
Then, we only need to construct the optimal policies satisfying Eq. (\ref{eqn_threshold}) for the other average power-delay pairs on curve $\mathcal{L}$.
In particular, we show the construction by using the properties of $\mathcal{L}$ in Theorem \ref{theorem_tradeoff_curve_policy_vertex} and the threshold-based structure for the optimal policies on the vertices.

For each power-delay pair $(P,D)$ on $\mathcal{L}$, we can find a pair of adjacent vertices $(\hat{P},\hat{D})$ and $(\tilde{P},\tilde{D})$, under which the power-delay pair is exactly on the line segment with the two vertices as the endpoints.
According to Theorem \ref{theorem_01}, we have that the pair of vertices on the curve $\mathcal{L}$ is generated by two threshold-based deterministic policies $\boldsymbol{\hat{F}}^\ast$ and $\boldsymbol{\tilde{F}}^\ast$, respectively. In other words, both the policies satisfy Eq. (\ref{eqn_deterministic_threshold}) as well as Eq. (\ref{eqn_threshold}).
Meanwhile, the two policies $\boldsymbol{\hat{F}}^\ast$ and $\boldsymbol{\tilde{F}}^\ast$ will employ different transmission rates only on a particular queue length and arrival rate.
As a result, according to Lemma \ref{lemma_linearcombination}, we can formulate the corresponding optimal policy for $(P,D)$ as the convex combination of the two threshold-based deterministic policies.%

Considering the two deterministic policies $\boldsymbol{\hat{F}}^\ast$ and $\boldsymbol{\tilde{F}}^\ast$ for the two adjacent vertices are threshold-based, we have that there exist the specific transmission rate $s^\ast$ and arrival rate $a^\ast$, under which the corresponding thresholds for the two policies are different.
Further, we have that the thresholds under the two policies are adjacent on the queue length, i.e., $|q_{\boldsymbol{\hat{F}}^\ast}(a^\ast,s^\ast)-q_{ \boldsymbol{\tilde{F}}^\ast}(a^\ast,s^\ast)|=1$.
Therefore, we show that the threshold-based optimal policy satisfies Eq. (\ref{eqn_threshold}) for each $(P,D)$ on curve $\mathcal{L}$, and the proof is completed.
\end{IEEEproof}

For a given threshold-based optimal policy $\boldsymbol{F}^\ast$, we obtain a series of thresholds $\{q_{\boldsymbol{F}^\ast}(s,a):~0\le{}s\le{}S\}$ under different arrival rates $a$.
Based on the order relation of $\{q_{\boldsymbol{F}^\ast}(s,a)\}$ in \mbox{Theorem \ref{theorem_threshold}}, we have that the transmission rate increases with the increase of the queue length.
Moreover, according to Theorem \ref{theorem_threshold}, the threshold-based optimal policy can be expressed as the convex combination of two adjacent deterministic threshold-based policies shown in Theorem \ref{theorem_01}.
As a result, for each system state $(q[n],a[n])$ except $(q_{\boldsymbol{F}^\ast}(s^\ast,a^\ast),a^\ast)$, we determine the transmission rates for AWGN channels by the queue length and arrival rate with the probability as $1$.
While the queue length is $q_{\boldsymbol{F}^\ast}(s^\ast,a^\ast)$ and arrival rate is $a^\ast$, the transmission rate is given as $s^\ast$ and $s^\ast-1$ with probabilities ${f^\ast}_{q_{\boldsymbol{F}^\ast}(s^\ast,a^\ast),a^\ast}^{s^\ast}$ and ${f^\ast}_{q_{\boldsymbol{F}^\ast}(s^\ast,a^\ast),a^\ast}^{s^\ast-1}$, respectively.%

\vspace{-5mm}
\subsection{Algorithm to Obtain the Optimal Tradeoff}
\vspace{-2mm}

\begin{figure}[!t]
	\centering
	\includegraphics[width=0.50\columnwidth,height=0.41\columnwidth]{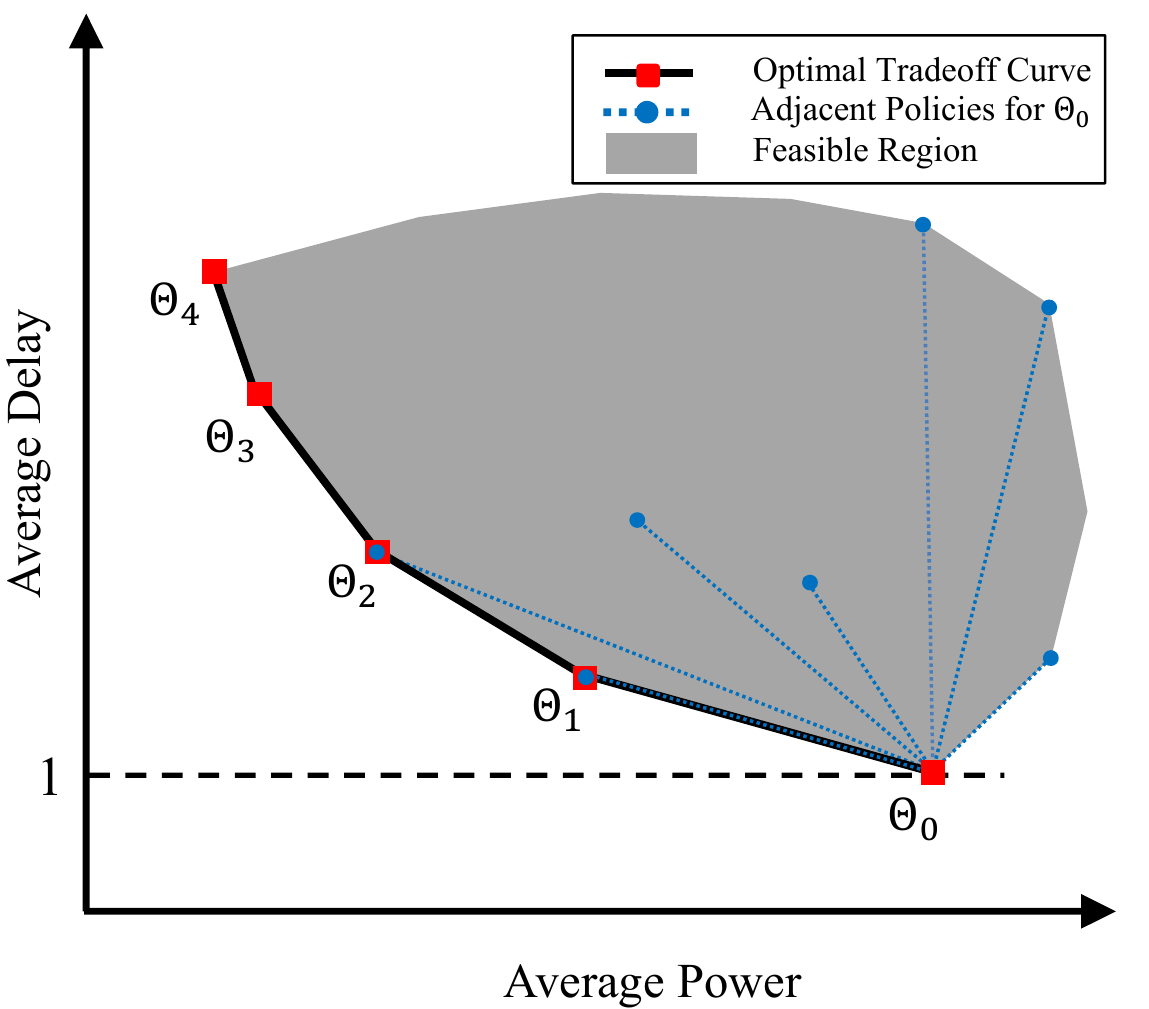}
	\vspace{-7mm}
	\caption{Demonstration of the algorithm to obtain the optimal delay-power tradeoff curve.}
	\label{fig_demo_algo}
	\vspace{-4mm}
\end{figure}

We finally develop a threshold-based algorithm to efficiently obtain the optimal delay-power tradeoff curve for AWGN channels.
In this way, the minimized delay can be generated by the optimal threshold-based policy for the given power constraint, which will be adjusted by practical systems based on the time varying delay and power efficiency requirements.
Considering the piecewise linearity of the optimal tradeoff curve $\mathcal{L}$, we first attain all the vertices of $\mathcal{L}$ and the corresponding threshold-based optimal deterministic policies.
As shown in \mbox{Algorithm \ref{algo_curve}}, we search the vertices sequence $\{\Theta_0,\Theta_1,\cdots,\Theta_N\}$ starting from $\Theta_0$ with an iteration procedure.
For the vertex $\Theta_0$ in \figurename~\ref{fig_demo_algo}, we obtain it by the policy that transmits the packets as soon as they arrive at the buffer.
In particularly, we denote this transmission policy by $\boldsymbol{F}_0$.

We next present the iteration procedure in Algorithm \ref{algo_curve} to find the current vertex $\Theta_{n+1}$ based on the previous vertex $\Theta_n$.
With the optimal deterministic policy $\boldsymbol{F}_{n}^\ast$ for previous vertex $\Theta_n=(\hat{P}_n,\hat{D}_n)$, we can detect the current vertex $\Theta_{n+1}=(\hat{P}_{n+1},\hat{D}_{n+1})$ by focusing on all the adjacent threshold-based deterministic policies of $\boldsymbol{F}_{n}^\ast$.
Overall the candidates of transmission policies, we obtain the threshold-based optimal deterministic policy $\boldsymbol{F}_{n+1}^\ast$ for vertex $\Theta_{n+1}$ based on the decreasing and convexity of $\mathcal{L}$.
More specifically, the average power-delay pair generated by $\boldsymbol{F}_{n+1}^\ast$, i.e., $\Theta_{n+1}$, has the slower increment of the average delay per decrement of the average power consumption starting from vertex $\Theta_n$ than that generated by any other candidate.
Therefore, the current vertex $\Theta_{n+1}$ and optimal policy $\boldsymbol{F}_{n+1}^\ast$ can be determined by enumerating all the deterministic policies that are adjacent with $\boldsymbol{F}_{n}^\ast$.
Further, we narrow down the alternatives of policy $\boldsymbol{F}_{n+1}^\ast$ by using the threshold-based structure presented in Theorem \ref{theorem_threshold}.
In Algorithm \ref{algo_curve}, we denote by $\mathcal{F}_p$ the set of the threshold-based policies under which the previous vertex $\Theta_n$ is generated as the average power-delay pair.
By enumerating the adjacent threshold-based deterministic policies for each policy in $\mathcal{F}_p$, we can obtain the current vertex $\Theta_{n+1}$ and the corresponding threshold-based optimal policies.
During the searching process, we backlog the candidate of the optimal policy in set $\mathcal{F}_c$, under which a less absolute slope and a lower power decreasing can be obtained on the power-delay plane.
As a result, when we traverse all the optimal policies that generate the vertex $\Theta_{n}$, the threshold-based optimal deterministic policy $\boldsymbol{F}_{n+1}^\ast$ is also attained for the current vertex $\Theta_{n+1}$.

\setlength{\textfloatsep}{5pt}
\begin{algorithm}[t]
		\setstretch{0.85}
		\caption{Obtain the Optimal Delay-Power Tradeoff for AWGN channels}
		\begin{algorithmic}[1]
			\State $\boldsymbol{F}\leftarrow{}\boldsymbol{F}_0$, $n\leftarrow{}0$
			\State $D_{\boldsymbol{F}}\leftarrow \text{average delay under policy}~ \boldsymbol{F}$, $P_{\boldsymbol{F}} \leftarrow \text{average power under policy}~ \boldsymbol{F} $
			\State $\mathcal{F}_c \gets [\boldsymbol{F}]$, $D_c \gets D_{\boldsymbol{F}}$, $P_c \gets D_{\boldsymbol{F}}$
			\While{$\mathcal{F}_c \neq \emptyset$}
			\State $\mathcal{F}_p \gets \text{the set containing an arbitrary policy in }\mathcal{F}_c$, $\mathcal{F}_c \gets \emptyset$, $\tilde{\mathcal{F}}_p\leftarrow{}\emptyset$
			\State $D_p \gets D_c$, $P_p \gets D_c$, $slope \gets +\infty$
			\While{$\mathcal{F}_p \neq \emptyset$}
			\State{$\boldsymbol{F}=\mathcal{F}_p\,\text{.\,pop(0)}$, $\tilde{\mathcal{F}}_p\,\text{.append}(\boldsymbol{F})$, $\hat{\mathcal{F}}_p\leftarrow{}\emptyset$}
			\State 
			\begin{tabular}{@{}rl}
				$\mathcal{F}(\boldsymbol{F})\leftarrow$&the set of all threshold-based deterministic policies satisfying Eq. (\ref{eqn_deterministic_threshold})\\
				& with
				the only one different threshold comparing with $\boldsymbol{F}$ 
			\end{tabular}
			\ForAll{$\boldsymbol{F}'\in\mathcal{F}(\boldsymbol{F})$}
			\State $D_{\boldsymbol{F}'}\leftarrow \text{average delay under}~ \boldsymbol{F}'$, $P_{\boldsymbol{F}'} \leftarrow \text{average power under}~ \boldsymbol{F}' $
			\If{$D_{\boldsymbol{F}'} = D_p$, $P_{\boldsymbol{F}'} = P_p$, and $\boldsymbol{F}'\notin\tilde{\mathcal{F}}_p$}
			\State {$\hat{\mathcal{F}}_p\,\text{.\,append}(\boldsymbol{F}')$}
			\ElsIf{$D_{\boldsymbol{F}'}=D_c$ and $P_{\boldsymbol{F}'}=P_c$}
			\State $\mathcal{F}_c\,\text{.\,append}(\boldsymbol{F}')$
			\ElsIf{$\frac{D_{\boldsymbol{F}'}-D_p}{P_p-P_{\boldsymbol{F}'}}<{}slope$ or $\frac{D_{\boldsymbol{F}'}-D_p}{P_p-P_{\boldsymbol{F}'}}={}slope,P_{\boldsymbol{F}'}>P_c$}
			\State $\mathcal{F}_c \gets [\boldsymbol{F}']$, $D_c \gets D_{\boldsymbol{F}'}$, $P_c \gets P_{\boldsymbol{F}'}$, $slope \gets \frac{D_{\boldsymbol{F}'}-D_p}{P_p-P_{\boldsymbol{F}'}}$
			\EndIf
			\EndFor
			\State {$\mathcal{F}_p\leftarrow\hat{\mathcal{F}}_p$}
			\EndWhile
			\State {$n\leftarrow{}n+1$}
			\State {$\Theta_{n}=(P_c,D_c)$, $\boldsymbol{F}^\ast_{n}=\mathcal{F}_{c}\,\text{. pop(0)}$}
			\EndWhile
		\end{algorithmic}
		\label{algo_curve}
	\end{algorithm} 

\setlength{\textfloatsep}{15pt}

Considering all the vertices are detected for curve $\mathcal{L}$, we finally show the optimal delay-power tradeoff under an arbitrary power constraint.
With the power constraint $P_{\rm{th}}$ given, we construct the corresponding optimal policy as a convex combination of two threshold-based policies.
According to Theorem \ref{theorem_threshold}, the two threshold-based policies corresponds to two adjacent vertices $(\hat{P}_n,\hat{D}_n)$ and $(\hat{P}_{n+1},\hat{D}_{n+1})$ that satisfy $(\hat{P}_{n}-P_{\text{th}})(\hat{P}_{n+1}-P_{\text{th}})\le{}0$.
In this way, we can find the two adjacent vertices on $\mathcal{L}$ by checking the sequence $\{\Theta_0,\cdots,\Theta_N\}$.
Considering the sequence is permuted with the power components increasing, we will end the research when finding the first vertex whose power component is less than $P_{\text{th}}$.
According to Lemma \ref{lemma_linearcombination}, we obtain the multiplier of the convex combination by the binary search over interval $[0,1]$.
By this means, the optimal delay-power tradeoff can be demonstrated under an arbitrary power constraint.
The threshold-based optimal transmission policy is also effectively formulated based on the threshold-based deterministic policies for the vertices.

Furthermore, we present the complexity of the proposed algorithm.
Considering an iteration process is employed for Algorithm 1, we first show the maximum number of iterations that search the adjacent policies for set $\mathcal{F}_p$; then analyze the complexity in each iteration.
As indicated in Algorithm 1, we update set $\mathcal{F}_p$ in each iteration by changing one particular state's transmission rate.
Meanwhile, under two arbitrary deterministic policies, the number of different transmission rates is no more than the number of system states, i.e., $QA$.
As a result, the number of iterations is no more than $QA$.
For each iteration, we further calculate the average delay and power for the $AS$ adjacent threshold-based policies of $\boldsymbol{F}$, where the most time-consuming operation for each candidate, that is the matrix inversion, costs $\mathcal{O}(Q^3A^3)$ in terms of time.
In this way, the time complexity of Algorithm 1 is $\mathcal{O}(Q^4A^5S)$.
Moreover, considering the set $\tilde{\mathcal{F}}_p$ has the most space consumption with the maximum number of policies as $QA$, we have that the space complexity is $\mathcal{O}(Q^2A^2S)$, where each policy is contained in $\tilde{\mathcal{F}}_p$ with the $QAS$ probabilities stored.

%
%

For practical systems, we can formulate a trajectory-sampling version of the algorithm.
More specifically, we generate the average delay and power as the mean value of $\frac{1}{\alpha}q[n]$ and $\rho[n]$ based on a long-term sampling of $s[n]$, $q[n]$, as well as $a[n]$.
The optimal delay-power tradeoff is then presented for the practical systems over AWGN channels without prior need of arrival statistics.

\vspace{-6mm}
\section{Optimal Delay-Power Tradeoff for Block Fading Channels}
\vspace{-4mm}

In this section, we extend the optimal delay-power tradeoff over block fading channels.
Based on the analyses of the optimal tradeoff for AWGN channels, we first show the optimal delay-power tradeoff for block fading channels by converting the CMDP to an LP problem.
By solving the equivalent LP problem, we then formulate an optimal delay-power tradeoff curve, where we show the properties for the curve that are same as those in Section III.
We finally present the optimal transmission policies over the fading channel with a threshold type of structure on the queue length.
For the optimal threshold-based policies, we further show an order relation of the thresholds under different channel states, when the power functions follow a particular condition.

First, we present the optimal delay-power tradeoff over block fading channels.
For the generalized system over the fading channel, we employ the steady-state analysis for each transmission policy $\boldsymbol{F}=\{f_{q,a,\iota}^s:\forall~q,~a,~\iota,~s\}$, as presented in Section III-A.
For this purpose, we formulate a Markov reward process for each given policy $\boldsymbol{F}$, through which the average delay and power consumption are presented by the steady-state probability.
In this way, we further show the optimal delay-power tradeoff by using an LP problem, where all the obtainable power-delay pairs are presented for the transmission policies in terms of the state-action frequencies $\{x_{q,a,\iota}^s\}$.
In particular, we present the LP problem as follows.
{\setlength\abovedisplayskip {2pt plus 7pt minus 8pt}
	\setlength\belowdisplayskip {2pt plus 7pt minus 8pt}
	\setlength\jot{-1pt}
\begin{subequations}
	\label{eqn_lp_fading}
	\begin{align}
	\min\limits_{\{x_{q,a,\iota}^{s}\}} \quad
	& \frac{1}{\alpha}\sum_{q=0}^{Q} \sum_{a=0}^{A} \sum_{\iota=1}^{L} \sum_{s=0}^S q x_{q,a,\iota}^{s} \label{obj_1_fading}\\
	\text{s.t.}\quad
	& \sum_{q=0}^{Q} \sum_{a=0}^{A} \sum_{\iota=1}^{L} \sum_{s=0}^S P_{h_\iota}(s) x_{q,a,\iota}^{s} \le P_{\text{th}} \label{con_P_fading}\\
	&\sum_{q=\max\{q'-a',0\}}^{\min\{q'-a'+S,Q\}}\sum_{a=0}^{A}\sum_{\iota=1}^{L}\sum_{s=0}^{S}\gamma_{a,a'}\eta_{\iota'}\notag\\
	&\quad x_{q,a,\iota}^{s}\mathbbm{1}_{\{s=q+a'-q'\}}=\sum_{s=0}^{S}x_{q',a',\iota'}^{s} \quad \forall~q',~a',~\iota' \label{con_1_fading}\\
	& \sum_{q=0}^{Q} \sum_{a=0}^{A} \sum_{\iota=1}^{L} \sum_{s=0}^S x_{q,a,\iota}^{s}=1 \label{con_2_fading}\\
	& x_{q,a,\iota}^s\ge 0 \qquad \forall~q,~a,~\iota,~s \label{con_3_fading}.
	\end{align}
\end{subequations}}%
By solving the LP problem under different power constraints $P_{\rm{th}}$, we then formulate the optimal delay-power tradeoff curve, which contains all the optimal average power-delay operating points under different power constraints.
With the same method in Sections III-B and III-C employed, we straightforwardly obtain the same properties of the optimal tradeoff curve as follows.
\vspace{-4mm}
\begin{theorem}\label{th_curve_fading}
	For block fading channels, the optimal delay-power tradeoff curve is piecewise linear, decreasing, and convex.
	The vertices of the optimal tradeoff curve are obtained by a series of deterministic transmission policies with unichains.
	For each two adjacent vertices, the corresponding two policies have different transmission rates only on one state.
\end{theorem}
\vspace{-4mm}
\begin{IEEEproof}
	The proof of this theorem is directly taken from the method of Theorems \ref{theorem_piecewise_linear} and \ref{theorem_tradeoff_curve_policy_vertex}.%
\end{IEEEproof}
\vspace{-2mm}%
As a result, the optimal delay-power tradeoff over the fading channel is obtained by solving the LP problem (\ref{eqn_lp_fading}), where the optimal policies are generated by the optimal solutions based on the extension of Eq. (\ref{x_to_F}) for fading channels.
By jointly exploiting the results in Theorem \ref{th_curve_fading} and Lemma \ref{lemma_linearcombination} over fading channels, we have that the optimal average delays are obtained by the corresponding optimal policies regardless of the initial system state.

Based on the analyses of the optimal tradeoff curve, we finally show that the optimal delay can be obtained by the optimal threshold-based policies over the fading channel.
With a similar way indicated in Section IV, we show the threshold-based structure of the optimal policies in the following theorem.
In particular, we first present the optimal deterministic threshold-based policies for the vertices of the tradeoff curve, where we employ the same method in Theorem \ref{theorem_01} for the CMDP generated over the fading channel.
Then, for other points on the optimal tradeoff curve, we show the threshold-based structure of the optimal policies by presenting them as the convex combination of two adjacent deterministic threshold-based policies, as indicated in Theorem \ref{theorem_threshold}.
We show the optimal threshold-based policies in the following theorem.%
\vspace{-5mm}
\begin{theorem}
	The optimal policy $\boldsymbol{F}^\ast$ exists $(A\!+\!1)\!\times\!(S\!+\!1)\!\times\!{}L$ thresholds $q_{\boldsymbol{F}^\ast}(s,a,\iota)$, \mbox{where we} have $0\!\le\!{}q_{\boldsymbol{F}^\ast}(0,\!a,\!\iota) \!\le\! \cdots \!\le\! q_{\boldsymbol{F}}(S,\!a,\!\iota)\!\le\!{}Q$ for each arrival rate $a,~0\le{}a\le{}A$ and index of channel state $\iota,~1\le{}\iota\le{}L$.
	With the thresholds $q_{\boldsymbol{F}^\ast}(s,a,\iota)$ given, the optimal policy $\boldsymbol{F}^\ast$ satisfies
	{\setlength\abovedisplayskip {2pt plus 5pt minus 6pt}
		\setlength\belowdisplayskip {2pt plus 5pt minus 6pt}
		\setlength\jot{0pt}
		\begin{equation}\label{eqn_threshold_fading}
		\left\{
		\begin{array}{ll}
		{f^\ast}_{q,a,\iota}^s=1 & q_{\boldsymbol{F}^\ast}(s\!-\!1,a,\iota)<q\le q_{\boldsymbol{F}^\ast}(s,a,\iota), a\!\neq\!a^\ast ~\text{or}~ s\!\neq\!s^\ast~\text{or}~\iota\!\neq\!\iota^\ast\\
		{f^\ast}_{q,a,\iota}^s=1 & q_{\boldsymbol{F}^\ast}(s\!-\!1,a,\iota)<q < q_{\boldsymbol{F}^\ast}(s,a,\iota),a\!=\!a^\ast ~\text{and}~ s\!=\! s^\ast~\text{and}~\iota\!=\!\iota^\ast\\
		{f^\ast}_{q,a,\iota}^s+{f^\ast}_{q,a,\iota}^{(s-1)}=1&q = q_{\boldsymbol{F}^\ast}(s,a,\iota),a= a^\ast ~\text{and}~ s= s^\ast~\text{and}~\iota=\iota^\ast\\
		{f^\ast}_{q,a,\iota}^s=0 & \text{otherwise}.
		\end{array}
		\right.
		\end{equation}}%
	where the specific $s^\ast$, $a^\ast$, and $\iota^\ast$ are given by $\boldsymbol{F}^\ast$, and $q_{\boldsymbol{F}^\ast}(-1,a,\iota)=-1$ for each $a$ and $\iota$.
	\label{theorem_threshold_fading}
\end{theorem}%
\vspace{-5mm}
\begin{IEEEproof}
	The proof of this theorem is directly taken from the method of Theorems \ref{theorem_01} and \ref{theorem_threshold}.%
\end{IEEEproof}

With the threshold-based structure of the optimal policies, we can efficiently determine the transmission rate for the adaptive transmitter over fading channels.
With the current arrival rate $a[n]$ and channel state $h[n]$ given as $a$ and $h_\iota$, respectively, we present the transmission rate by comparing the current queue length with the series of thresholds $\{q_{\boldsymbol{F}^\ast}(s,a,\iota):0\le{}s\le{}S\}$.
As a result, we can also obtain the optimal delay-power tradeoff for fading channels by developing a similar algorithm as Algorithm \ref{algo_curve}.
Moreover, we show an order relation of the thresholds under different transmission rate in the following theorem.
\vspace{-5mm}
\begin{theorem}\label{theorem_last}
	The thresholds $q_{\boldsymbol{F}^\ast}(s,a,\iota),~1\le{}\iota\le{}L$ of the optimal policy $\boldsymbol{F}^\ast$ satisfy 
	{\setlength\abovedisplayskip {2pt plus 5pt minus 6pt}
		\setlength\belowdisplayskip {2pt plus 5pt minus 6pt}
		\vspace{-3mm}
	\begin{equation}\label{threshold_order}
		q_{\boldsymbol{F}^\ast}(s,a,\iota^+)\le{}q_{\boldsymbol{F}^\ast}(s,a,\iota^-),\quad \forall~ 1\le{}\iota^-<\iota^+\le{}L,
		\vspace{-3mm}
	\end{equation}}%
	for each transmission rate $s$ and arrival rate $a$, 
	when power functions $P_{h_\iota}(s),~1\!\le{}\!\iota\!\le{}\!L$ satisfies%
		\vspace{-3mm}
	\begin{equation}\label{power_condition}
		P_{h_{\iota^+}}(s^+)-P_{h_{\iota^+}}(s^-)\le{}P_{h_{\iota^-}}(s^+)-P_{h_{\iota^-}}(s^-),
		\vspace{-3mm}
	\end{equation}%
	where we have $0\le{}s^-<{}s^+\le{}S$.
\end{theorem}
\vspace{-5mm}
\begin{IEEEproof}
See Appendix \ref{Appendix_D}.
\end{IEEEproof}
\vspace{-2mm}

According to the order relation of thresholds in Eq. (\ref{threshold_order}), a greater rate will employed for a better channel condition under the optimal threshold-based policies, if the condition in Eq. (\ref{power_condition}).
Actually, for a typical communication system, we have that the power consumption for a transmission rate is inversely proportional to the square of amplitude of channel coefficient, i.e., $\frac{P_{h_{\iota^+}}(s)}{|h_{\iota^-}|^2}=\frac{P_{h_{\iota^-}}(s)}{|h_{\iota^+}|^2}$.
As a result, we can straightforwardly check the condition in Eq. (\ref{power_condition}) in the typical system, through which the order relation of thresholds of optimal policies is satisfied.

\vspace{-6mm}
\section{Numerical Results}
\vspace{-4mm}

In this section, we present the numerical results to validate the optimal delay-power tradeoff for the adaptive transmitter with Markov random arrivals.
In a practical scenario, we consider that the maximum transmission rate $S$ is equal to $3$, under which we employ three optional modulations BPSK, QPSK, or 8-PSK to transmit 1,~2,~or~3 packets in a timeslot, respectively.
We assume that each packet contains 10,000 bits and time duration of timeslot is 10 ms.
With the bandwidth as 1 MHz and the one-sided noise power spectral density $N_0$ as $-150$ dBm/Hz, we calculate the transmission powers over AWGN channels as $P(0)=0$ W, $P(1)=9.0\times 10^{-12}$ W, $P(2)=18.2\times 10^{-12}$ W, and $P(3)=59.5\times 10^{-12}$ W, by which the bit error rate as $10^{-5}$ is provided.
Moreover, we consider a specific class of the arrival processes.
For each arrival process, we determine the transition matrix $\boldsymbol{\Gamma}=[\gamma_{a,a'}]$ by a constant $\psi$ and a vector $\boldsymbol{\zeta}=[\zeta_0,\zeta_1,\cdots,\zeta_A]^T$.
In particular, we define matrix $\boldsymbol{\Gamma}$ by presenting each element $\gamma_{a,a'}$ as%
{\setlength\jot{1pt}
	\vspace{-2mm}
\begin{equation}
	\gamma_{a,a'} = \frac{1-\zeta_a}{A}+\frac{(A+1)\zeta_a-1}{A}\mathbbm{1}_{\{a'=(a+\psi)\!\!\!\!\mod (A+1)\}}.
	\vspace{-2mm}
\end{equation}}%
As a result, we construct an arrival process by using a tuple $(\boldsymbol{\zeta},\psi)$, and have that $\zeta_a\in[0,1]$ and $\psi\in\{-A,-A+1,\cdots,A\}$.

\begin{figure}[t] 
	\centering
	\includegraphics[width=0.68\columnwidth,height=0.48\columnwidth]{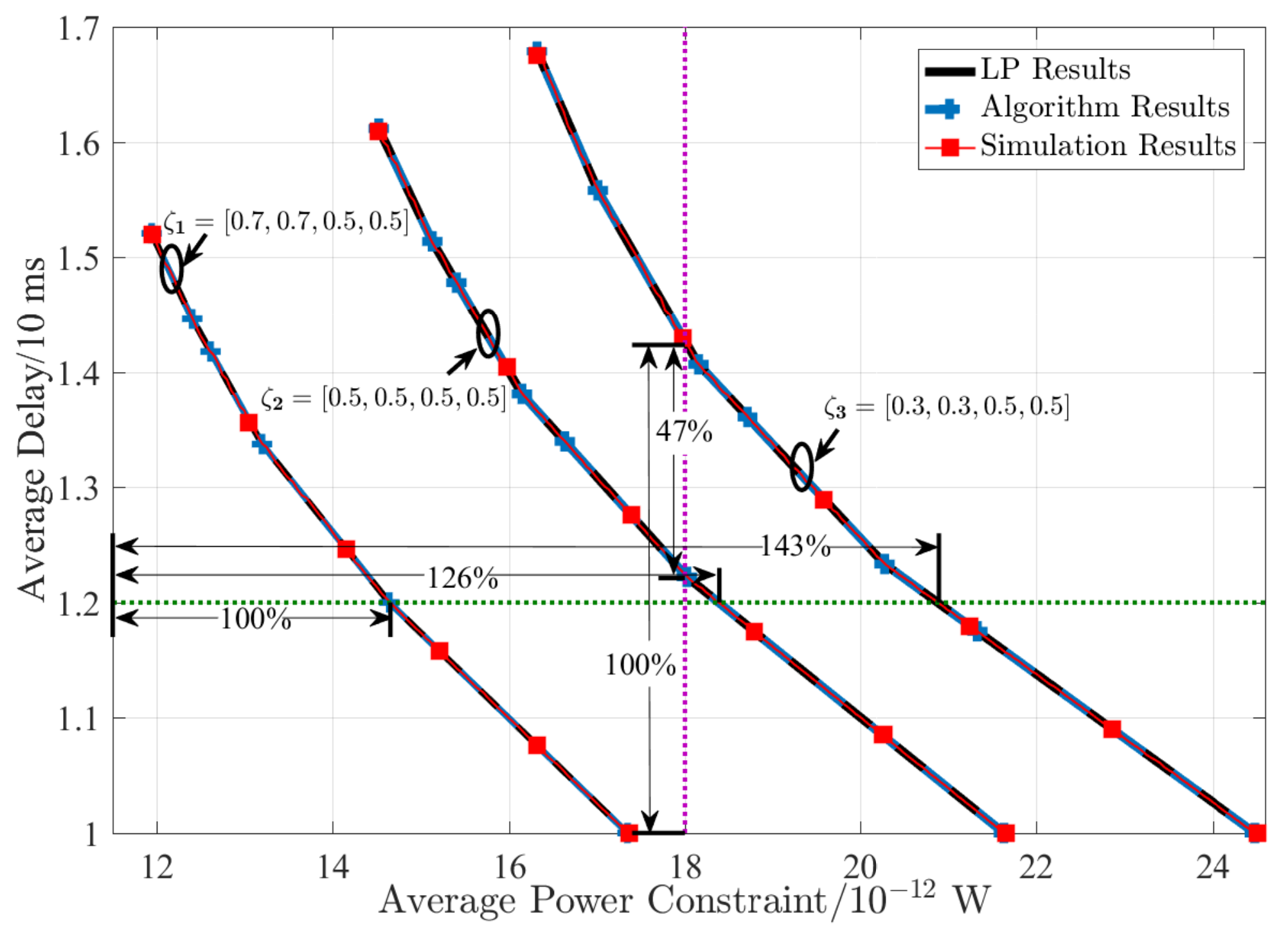}
	\vspace{-8mm}
	\caption{Optimal Delay-Power Tradeoff Curves}
	\label{fig_optimization_0}
	\vspace{-4mm}
\end{figure}

First, \figurename~\ref{fig_optimization_0} presents the optimal delay-power tradeoff curves for AWGN channels, where we consider the impact of different average arrival rates.
For the optimal tradeoff curves, we validate the theoretical results by using the Monte-Carlo simulation.
We assume the maximum arrival rate and transmission rate as $3$, and the buffer size as $7$.
The optimal delay-power tradeoff curves are next presented for the three different arrival processes, all of which are charactered as $(\boldsymbol{\zeta}_i,0),~i=1,2,3$.
In particular, we have $\boldsymbol{\zeta}_1=[0.7,0.7,0.5,0.5]$, $\boldsymbol{\zeta}_2=[0.5,0.5,0.5,0.5]$\mbox{, and} $\boldsymbol{\zeta}_3=[0.3,0.3,0.5,0.5]$.
The average rates for the three arrival processes are equal to $1.25,~1.50$, and $1.67$, respectively.
As presented in \figurename~\ref{fig_optimization_0}, the optimal delay-power tradeoff given by Algorithm \ref{algo_curve} and solving the LP problem can perfectly match the results that are given by the Monte-Carlo simulation.
In each optimal tradeoff curve, the optimal average delay is decreasing with the increase of average power consumption.
Further, a close observation shows that each curve is piecewise linear and convex, by which we confirm Theorem \ref{theorem_piecewise_linear}.
Then, we present different optimal delay-power tradeoff curves under different average arrival rates.
When the power constraint is \mbox{$P_{\text{th}}=18\times10^{-12}$ W,} the average delay under $(\boldsymbol{\zeta}_2,0)$ can reduce by $47\%$ compared with that under $(\boldsymbol{\zeta}_3,0)$.
To achieve the average delay $\mathsf{D}=1.2\times{}10$ms, arrival processes $(\boldsymbol{\zeta}_2,0)$ and $(\boldsymbol{\zeta}_3,0)$ require greater power consumptions, which are $126\%$ and $143\%$ of that \mbox{for $(\boldsymbol{\zeta}_1,0)$.}

\begin{figure}[!t]
	\centering
	\subfigure[The average transmission rates for the optimal policy~]
	{\includegraphics[width=0.45\columnwidth,height=0.26\columnwidth]{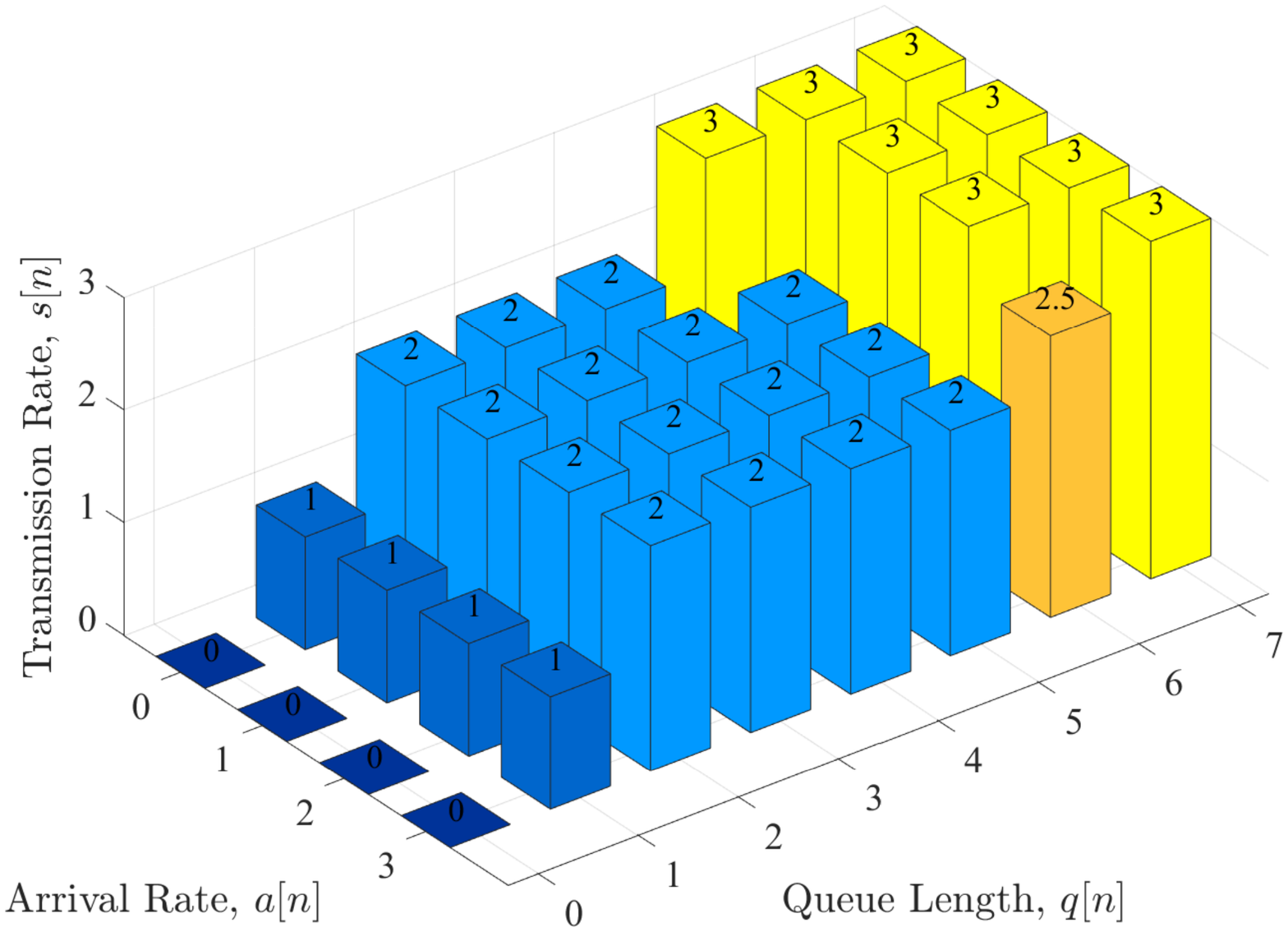}
		\label{fig_typical_1}}
	\subfigure[The thresholds for the optimal policy]{\includegraphics[width=0.45\columnwidth,height=0.26\columnwidth]{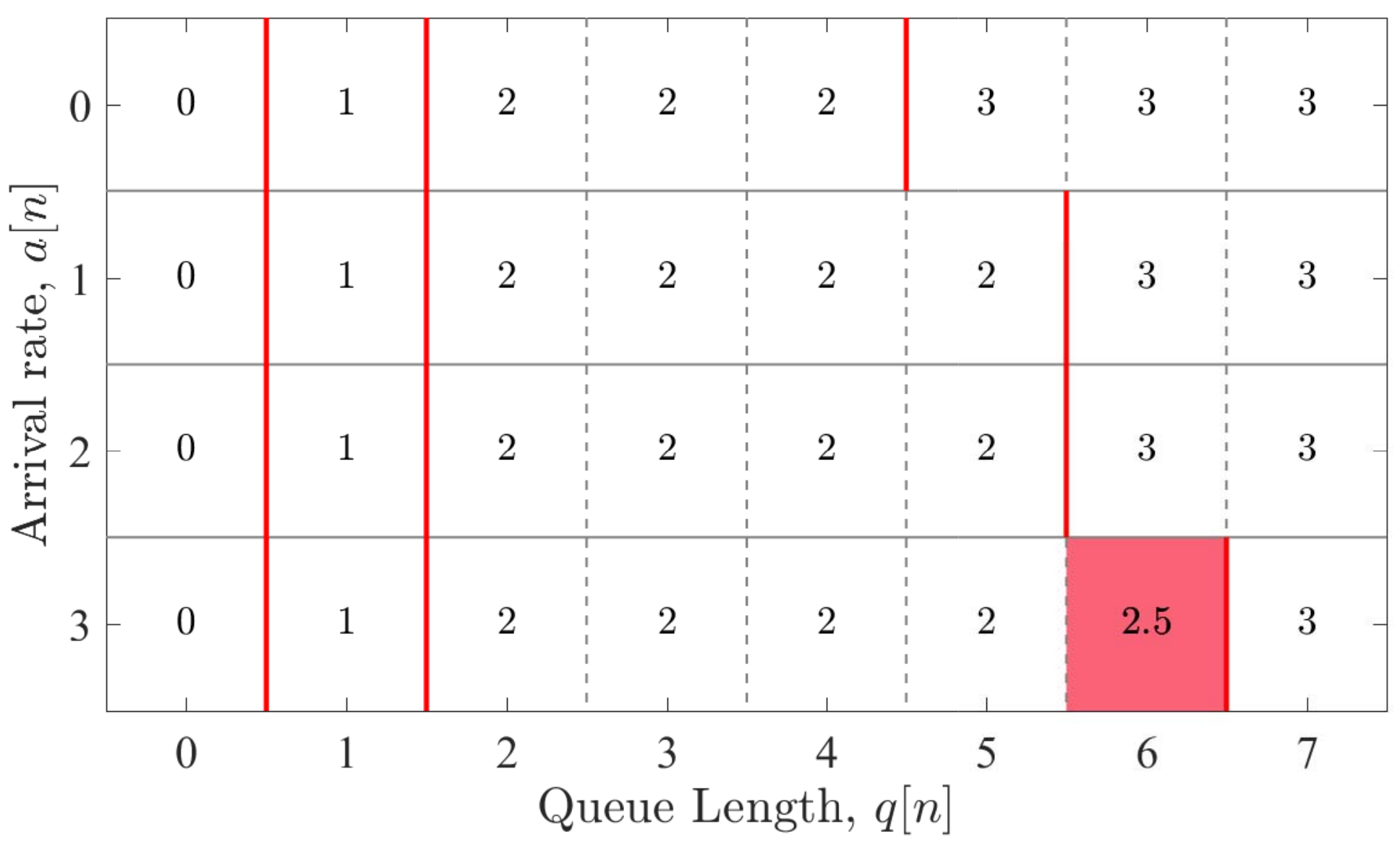}
		\label{fig_typical_2}}
	\vspace{-4mm}
	\caption{Typical optimal threshold-based policy}
		\vspace{-4mm}
	\label{fig_typical}
\end{figure}

Then, we turn our attention to the threshold-based structure of the optimal cross-layer transmission \mbox{policy} over AWGN channels.
\figurename~\ref{fig_typical} presents the typical threshold-based optimal policy $\boldsymbol{F}^\ast$ for the identified system configuration as \figurename~\ref{fig_optimization_0} with the arrival process given as $(\boldsymbol{\zeta}_2,0)$ and \mbox{ $P_{\text{th}}=14.82\times{}10^{-12}$ W.}
In \figurename~\ref{fig_typical_1}, we particularly show the average transmission rates under different queue lengths and arrival rates.
We also indicate the threshold-based structure in \figurename~\ref{fig_typical_2}, where we present the thresholds by red solid lines.
According to the order relation of the thresholds in Theorem \ref{theorem_threshold}, we show a greater transmission rate for a longer queue length under the optimal policy $\boldsymbol{F}^\ast$.
Following Theorem \ref{theorem_threshold}, we further present the typical policy as a convex combination of two adjacent deterministic policies, both of which exist a \mbox{threshold-based structure} in Theorem \ref{theorem_01}.
As a result, the transmission rates under $\boldsymbol{F}^\ast$ are deterministic for the system states except for a specific one with arrival rate and queue length as $3$ and $6$, respectively.


We next show the impact of different patterns of Markov arrivals to the optimal delay-power tradeoff over AWGN channels even if we employ the same average rate and covariance in these random arrivals.
In particular, we focus on the three arrival patterns that are denoted by $\mathcal{A}_1$, $\mathcal{A}_2$, and $\mathcal{A}_3$, the transition matrices of which are given by $(\kappa_1\boldsymbol{1},1)$, $(\kappa_2\boldsymbol{1},0)$, and $(\kappa_3\boldsymbol{1},-1)$, respectively.
We have $\kappa_i\in[0,1]$ for $i=1,2,3$, and all the elements of vector $\boldsymbol{1}$ are equal to $1$.
Then, the random arrivals under all the three patterns have the same steady-state probability distribution, and the steady-state probabilities of all the arrival rates are the same, \mbox{i.e., $\frac{1}{A+1}$.}
Therefore, the average arrival rate of each arrival process is equal to $\frac{A}{2}$.
To obtain the same covariances $\text{cov}(a[n],a[n+1])$ for the random arrivals under three different arrival patterns, we set $\kappa_1=\kappa_3=\kappa$ and $\kappa_2 = \frac{3-2\kappa}{10}$, under which we have $\text{cov}(a[n],a[n+1])=\frac{1-4\kappa}{10}$.

\begin{figure}[!t] 
	\centering
	\includegraphics[width=0.68\columnwidth,height=0.48\columnwidth]{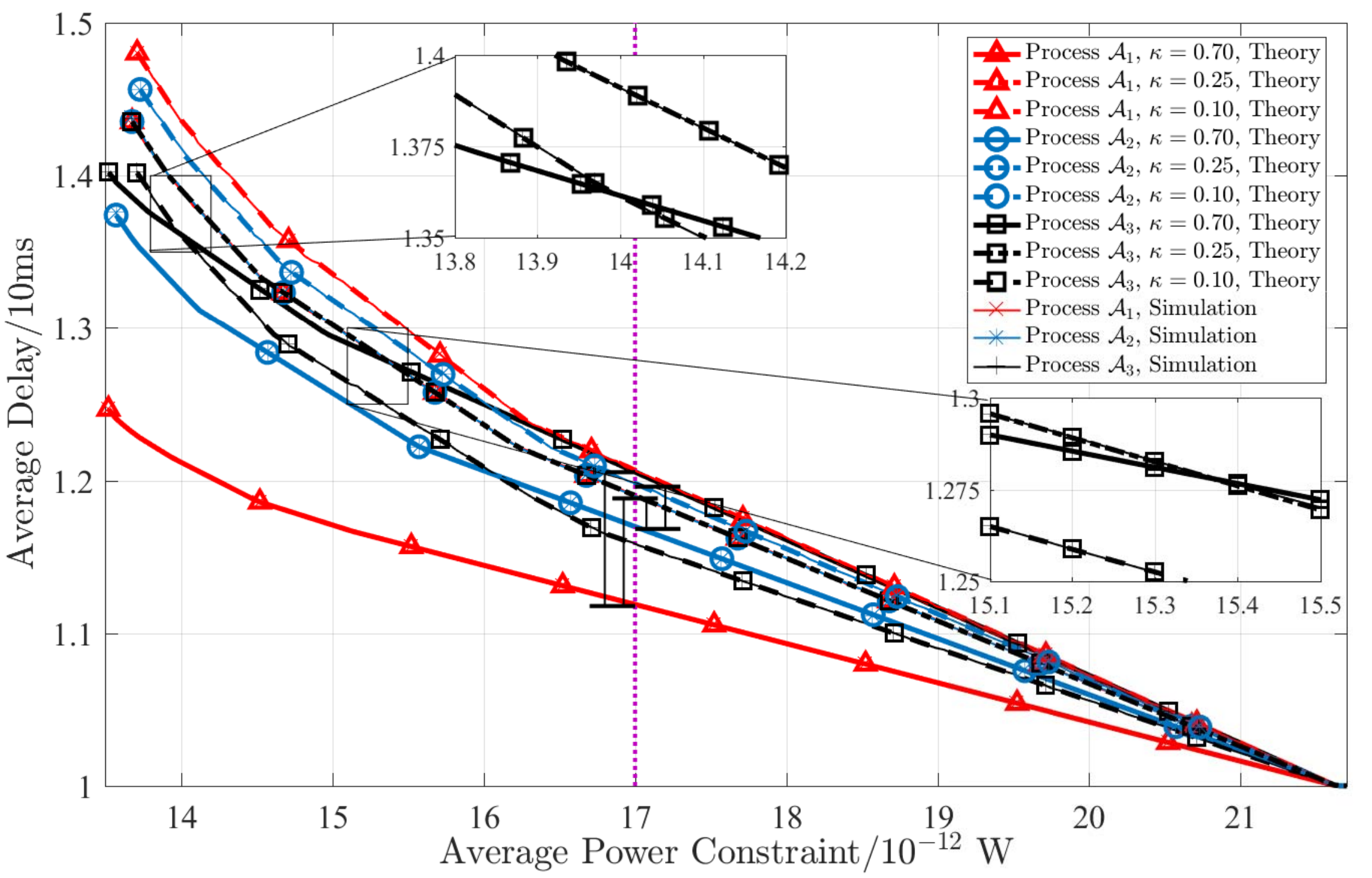}
	\vspace{-8mm}
	\caption{Optimal Delay-Power Tradeoff Curves under different arrival processes}
	\label{fig_arrival_pattern}
	\vspace{-4mm}
\end{figure}

As shown in \figurename~\ref{fig_arrival_pattern}, we present the optimal delay-power tradeoff curves for the three arrival patterns with parameter $\kappa$ given as $0.1,$ $0.25$, and $0.7$\footnote{When $\kappa$ is equal to $0.25$, we have the same arrival processes under the three arrival patterns, through which the corresponding curves are coincident}.
In particular, we have a lower average delay for arrival processes $\mathcal{A}_1$ and $\mathcal{A}_2$ if we increase $\kappa$, i.e., decrease covariance.
When $P_{\text{th}}=17\times10^{-12}$ W, the average delay under $\mathcal{A}_1$ with $\kappa=0.7$ can be reduced by $37\%$ and $43\%$ compared to that with $\kappa=0.25$ and $\kappa=0.1$.
As for arrival process $\mathcal{A}_2$, we have that the average delay is reduced by $11\%$ and $14\%$.
However, for arrival process $\mathcal{A}_3$, the average delays under the three value of $\kappa$ have different order relations with the varying of the average power constraint.

\begin{figure}[!t] 
	\centering
	\includegraphics[width=0.68\columnwidth,height=0.48\columnwidth]{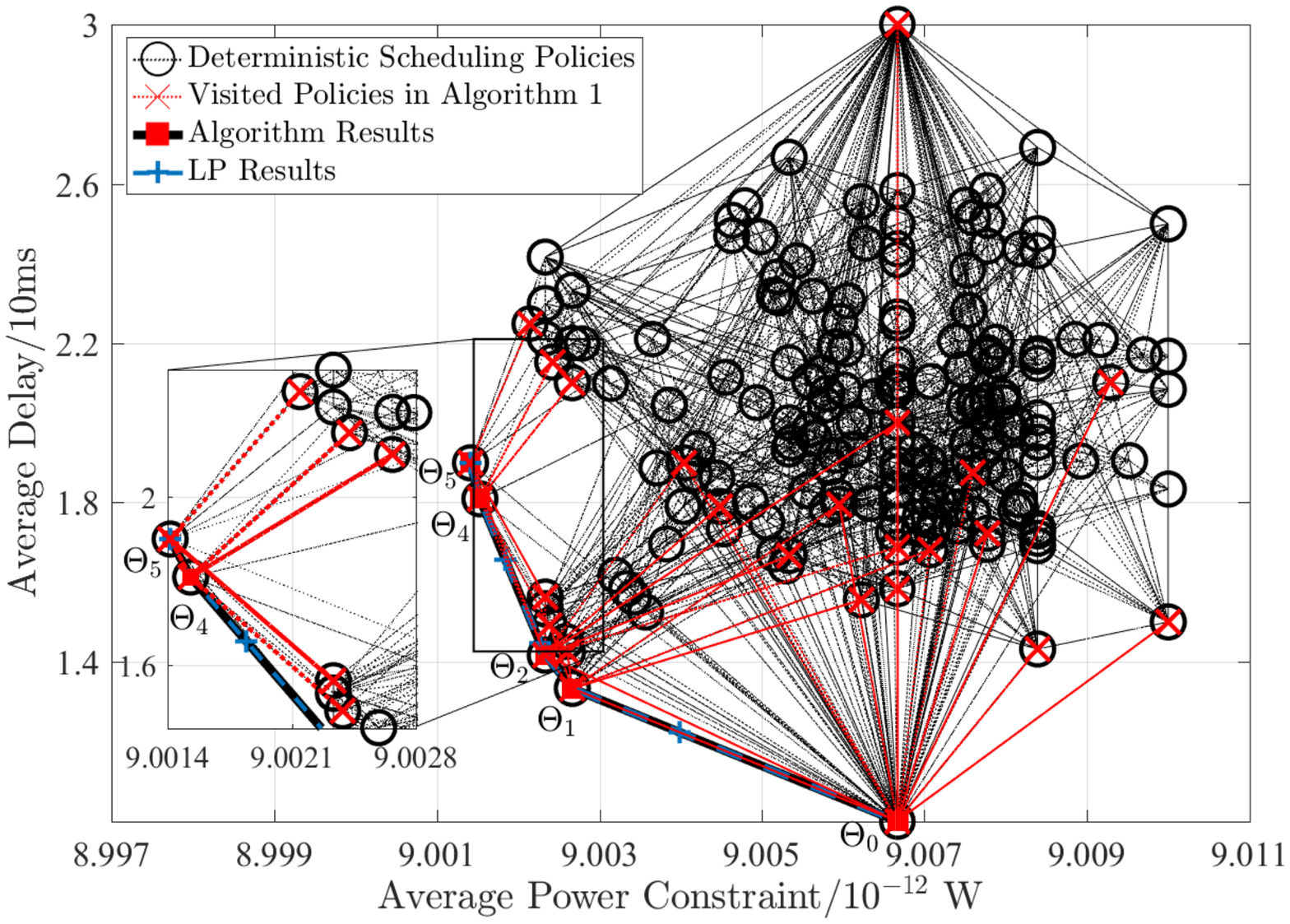}
	\vspace{-7mm}
	\caption{Optimal Delay-Power Tradeoff Curves}
	\label{fig_vertex}
	\vspace{-3mm}
\end{figure}

In \figurename~\ref{fig_vertex}, we present the procedure to obtain the optimal delay-power tradeoff for AWGN channels, which is given by Algorithm \ref{algo_curve}.
To simplify the figure, we assume $Q=4$ and $S=A=2$, and consider the arrival process presented by $([0.6,0.6]^T,1)$.
We first show the power-delay pairs obtained by the deterministic policies by using marker 'o'.
Further, we connect the two points generated by two adjacent policies by the black dash lines.
With the vertex $\Theta_0$ and corresponding policy $\boldsymbol{F}_0$ given, 
we seek the vertices $\Theta_{n}$ among the threshold-based deterministic policies that are adjacent with the previous optimal policies in set $\mathcal{F}_{p}$.
To present those investigated policies in Algorithm \ref{algo_curve}, we particularly show the corresponding power-delay pairs by marker '$\times$' and connect them with adjacent vertices on the optimal tradeoff curve $\mathcal{L}$ by the red dash lines. 
As shown in \figurename~\ref{fig_vertex}, the optimal delay-power tradeoff can be effectively obtained by Algorithm \ref{algo_curve}, where a few adaptive transmission policies are investigated over all the deterministic policies.

\begin{figure}[!t]
	\centering
	\subfigure[The optimal delay-power tradeoff curve over the block fading channel]
	{\includegraphics[width=0.45\columnwidth,height=0.27\columnwidth]{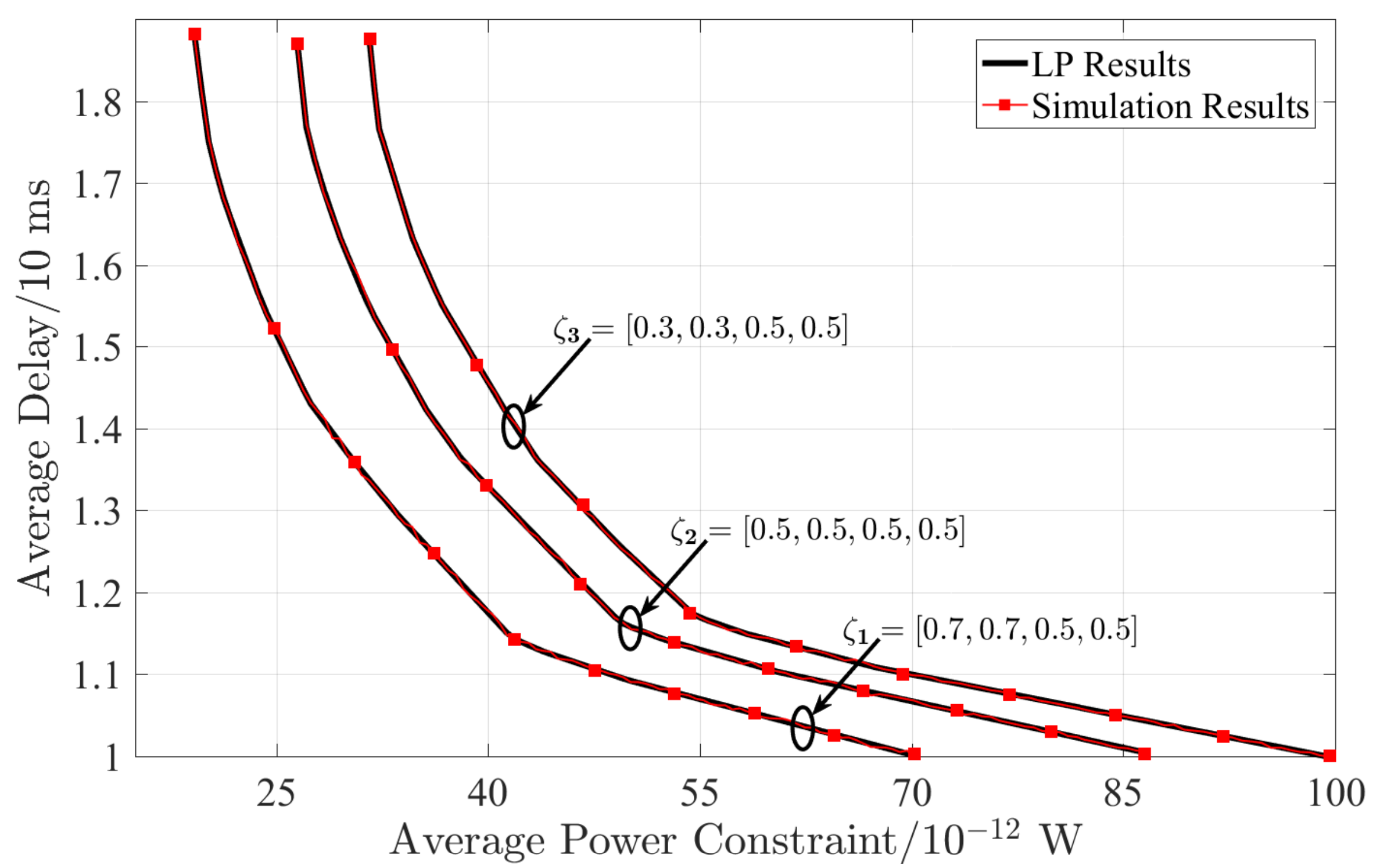}\label{curve_fading}}
	\subfigure[The average transmissin rate under the optimal policy with the arrival rate given as 2 ]{\includegraphics[width=0.45\columnwidth,height=0.27\columnwidth]{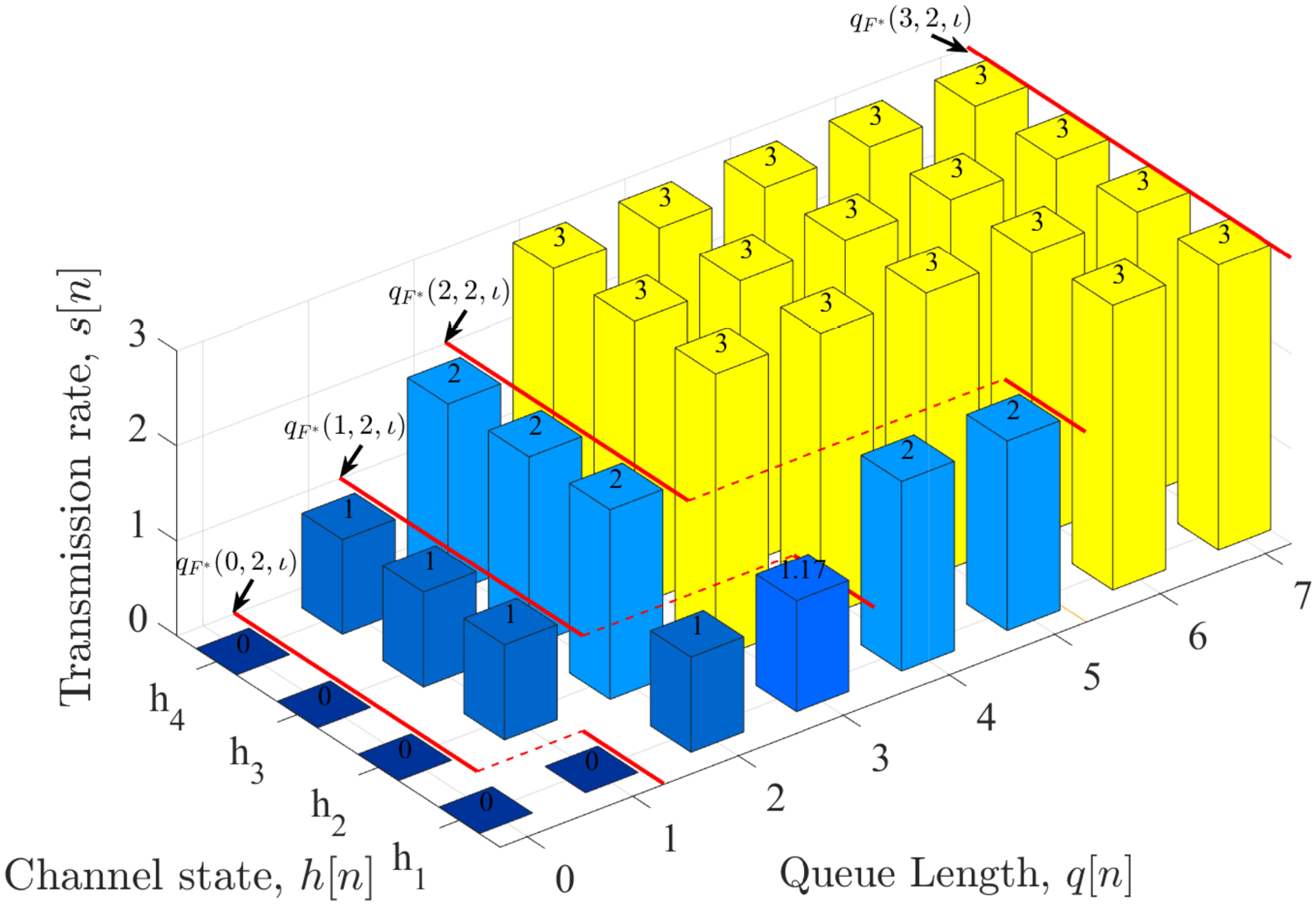}\label{policy_fading}}
	\vspace{-4mm}
	\caption{Optimal delay-power tradeoff over fading channel}
	\label{DP_fading}
	\vspace{-2mm}
\end{figure}

We finally show the optimal delay-power tradeoff for block fading channels.
In Fig. \ref{DP_fading}, we consider an $L$-state block fading channel with $L$ given as $4$.
In particular, for the fading channel, the amplitudes of the four channel states, i.e., $|h_\iota|,~\iota=1,\cdots,4$, are given as $0.314$, $2.50$, $3.54$, and $5.00$, respectively.
The corresponding probabilities $\eta_\iota$ are presented as $0.394$, $0.232$, $0.239$, and $0.135$.
We obtain the power consumptions under different channel states by define the power consumption function $P_{h_\iota}(s)$ as $\frac{P(s)}{|h_\iota|^2}$ for each $s$ and $\iota$.
As a result, we present the optimal delay-power tradeoff curves for different arrival processes in Fig. \ref{curve_fading}, where we employed the same system configuration as that Fig. \ref{fig_optimization_0}.
Moreover, we also show the average transmission rates under an optimal threshold-based policy in Fig. \ref{policy_fading} with the current arrival rate $a[n]$ given as $2$.
As indicated in Fig. \ref{policy_fading}, we present the threshold-based structure for the optimal policy, in which we show the thresholds on the queue lengths by red solid lines.
In Fig. \ref{policy_fading}, we further illustrate the order relation of the thresholds under different channel states that is given by Theorem \ref{theorem_last}.
By this means, under the current queue length, a greater transmission rate is employed for a better channel condition.

\vspace{-5mm}
\section{Conclusion}
\vspace{-3mm}

In this paper, we have obtained the optimal delay-power tradeoff required for transmission over a wireless link under Markov arrivals.
The problem can be formulated as a CMDP, under which we jointly consider the queue length, arrival rate, and channel state to minimize the average delay under an average power constraint.
To obtain the optimal delay-power tradeoff, we have shown an equivalent LP problem based on the steady-state analysis of the Markov reward process.
Varying the power constraints in the derived LP problem, we show that the optimal delay-power tradeoff curve is decreasing, convex and piecewise linear.
Based on these geometric properties, we have also presented the optimal adaptive transmission policies for the optimal power-delay pairs on the tradeoff curve.
Further, the threshold-based structure of the optimal policies has been demonstrated in the queue length by using the Lagrangian relaxation.
With the threshold-based structure, we have developed a threshold-based algorithm to efficiently obtain the optimal delay-power tradeoff for practical communications.

\appendices
\vspace{-4mm}
\section{Proof of Theorem \ref{theorem_tradeoff_curve_policy_vertex}}
\label{Appendix_A}
\vspace{-3mm}

The proof falls into three parts.
We first show that the policies generating the vertices of curve $\mathcal{L}$ with unichains.
To obtain a contradiction, we suppose that there exists a policy $\boldsymbol{F}$ for a vertex of $\mathcal{L}$, under which a multichain is generated with the number of closed classes as $I$.
Then, the set $\{x_{q,a}^s:\pi_{\boldsymbol{F}}(q,a)f_{q,a}^{s}\}$ is varied with the initial state.
Moreover, we can construct a series of policies $\hat{\boldsymbol{F}}_i,~i=1,2,\cdots,I$ with unichains, among which the policy $\hat{\boldsymbol{F}}_i$ employs the same transmission rates as policy $\boldsymbol{F}$ for each state in the $i$th recurrent closed class.
The existence of policy $\hat{\boldsymbol{F}}_i$ is provided by the communicating property of the CMDP in \cite[Section 8.3.1]{puterman2014markov}.
As a result, the same steady-state distribution is obtained under the policies $\hat{\boldsymbol{F}}_{i}$ and $\boldsymbol{F}$ with the system starting from the $i$th recurrent closed class.
In this way, the state-action frequencies $\{x_{q,a}^s\}$ generated by $\boldsymbol{F}$ can be expressed as the convex combination of state-action frequencies of policies $\hat{\boldsymbol{F}}_i$, where the convex multipliers are determined based on the initial state.
As a corollary, $\{x_{q,a}^s\}$ is not the vertex of $\mathcal{G}$.
Since set $\mathcal{R}$ is the projection of $\mathcal{G}$ and contains $\mathcal{L}$, we have that the vertices of $\mathcal{L}$ must be projected by the vertices of $\mathcal{G}$, which induces to a contradiction.

Then, we show that all the vertices of curve $\mathcal{L}$ are obtained by the deterministic policies.
For this purpose, we apply the similar consideration as \cite[Theorem 4.2]{altman1999constrained}.
This theorem shows that the vertices of $\mathcal{G}$ are generated by the deterministic policies, if all the considered policies have unichains.
With the above analysis, the proof is straightforwardly checked based on the theorem.

We finally show the relationship of policies for two adjacent vertices on $\mathcal{L}$.
We start the proof with the observation that the edge connecting the two adjacent vertices on curve $\mathcal{L}$ is the projection of an edge on $\mathcal{G}$, where the vertices are generated by the deterministic policies.
For the two adjacent vertices on $\mathcal{G}$, the corresponding deterministic policies are different only on one state.
The conclusion also holds for the degenerated case that the edges connecting a series of adjacent vertices of $\mathcal{L}$ are collinear.
In this way, the proof of this theorem is completed.

\vspace{-5mm}
\section{Proof of Lemma \ref{lemma_linearcombination}}
\label{Appendix_C}
\vspace{-3mm}

We begin by recalling that the probabilities of the transmission rates in policy $\boldsymbol{F}$ are the same as that in policies $\boldsymbol{F}'$ and $\boldsymbol{F}''$ for all the states except state $(\tilde{q},\tilde{a})$.
When the system state is given as $(\tilde{q},\tilde{a})$, we next randomly determine the employed policy as $\boldsymbol{F}'$ or $\boldsymbol{F}''$ with probabilities $\epsilon$ or $1-\epsilon$, respectively.
Considering that $\boldsymbol{F}'$ and $\boldsymbol{F}''$ have unichains, we can visit $(\tilde{q},\tilde{a})$ within a finite time duration starting from any other state under $\boldsymbol{F}'$ and $\boldsymbol{F}''$.
As a result, we can also obtain the identified random process under policy $\boldsymbol{F}$, i.e., the system visits $(\tilde{q},\tilde{a})$ starting from a given state.
Therefore, there exists only one recurrent closed class in the Markov chain under policy $\boldsymbol{F}$, i.e., policy $\boldsymbol{F}$ has a unichain.

Then, we present that the average power and delay under $\boldsymbol{F}$ is formulated as the convex combination of those under $\boldsymbol{F}'$ and $\boldsymbol{F}''$ based on the relationship of the three policies' state-action frequencies.
To this end, we first present $x_{q,a}^{s}$ under $\boldsymbol{F}$ by the method in \mbox{\cite[Eq. 4.3.8]{kao1997introduction} as} $
	x_{q,a}^{s}=\pi_{\boldsymbol{F}}(q,a)f_{q,a}^s=\frac{\mathbb{E}^{\boldsymbol{F}}_{\tilde{q},\tilde{a}}\{N(q,a,s)\}}{\mathbb{E}^{\boldsymbol{F}}_{\tilde{q},\tilde{a}}\{T\}},
$
where we have $T=\inf\{n\ge{}1:q[n]=\tilde{q},a[n]=\tilde{a}\}$ and $N(q,a,s)=\sum_{n=0}^{T-1}\mathbbm{1}_{\{q[n]=q,a[n]=a,s[n]=s\}}$.

Considering that the average power and delay in Eqs. (\ref{obj_1}) and (\ref{con_P}) are linear functions of $x_{q,a}^s$, we only need show the relationship of $x_{q,a}^s$, ${x'}_{q,a}^s$, and ${x''}_{q,a}^s$, where we define ${x'}_{q,a}^s=\pi_{\boldsymbol{F}'}(q,a){f'}_{q,a}^s$ and ${x''}_{q,a}^s=\pi_{\boldsymbol{F}''}(q,a){f''}_{q,a}^s$.
Based on the above definition of $x_{q,a}^s$, we have
\begin{align}
	x_{q,a}^{s}
	&=\frac{\epsilon\mathbb{E}^{\boldsymbol{F}'}_{\tilde{q},\tilde{a}}\{N(q,a,s)\}+(1-\epsilon)\mathbb{E}^{\boldsymbol{F}''}_{\tilde{q},\tilde{a}}\{N(q,a,s)\}}{\epsilon\mathbb{E}^{\boldsymbol{F}'}_{\tilde{q},\tilde{a}}\{T\}+(1-\epsilon)\mathbb{E}^{\boldsymbol{F}''}_{\tilde{q},\tilde{a}}\{T\}}\nonumber\\
	&=\frac{\epsilon\mathbb{E}^{\boldsymbol{F}'}_{\tilde{q},\tilde{a}}\{T\}{x'}_{q,a}^s+(1-\epsilon)\mathbb{E}^{\boldsymbol{F}''}_{\tilde{q},\tilde{a}}\{T\}{x''}_{q,a}^s}{\epsilon\mathbb{E}^{\boldsymbol{F}'}_{\tilde{q},\tilde{a}}\{T\}+(1-\epsilon)\mathbb{E}^{\boldsymbol{F}''}_{\tilde{q},\tilde{a}}\{T\}},
\end{align}%
where the first equality holds based on the above analysis of transmission process under $\boldsymbol{F}$.

By defining $\epsilon'=\frac{\epsilon\mathbb{E}^{\boldsymbol{F}'}_{\tilde{q},\tilde{a}}\{T\}}{\epsilon\mathbb{E}^{\boldsymbol{F}'}_{\tilde{q},\tilde{a}}\{T\}+(1-\epsilon)\mathbb{E}^{\boldsymbol{F}''}_{\tilde{q},\tilde{a}}\{T\}}$, we finally have $x_{q,a}^{s}=\epsilon'{x'}_{q,a}^s+(1-\epsilon'){x''}_{q,a}^s$.
Therefore, the average power and delay under policy $\boldsymbol{F}$ are given as $P_{\boldsymbol{F}}\!=\!(1-\epsilon')P_{\boldsymbol{F}'}\!+\!\epsilon' P_{\boldsymbol{F}''}$ and $D_{\boldsymbol{F}}\!=\!(1-\epsilon')D_{\boldsymbol{F}'}\!+\!\epsilon' D_{\boldsymbol{F}''}$, respectively.
An easy computation shows that $\epsilon'$ is monotone increasing with $\epsilon$ under the given $\mathbb{E}^{\boldsymbol{F}'}_{\tilde{q},\tilde{a}}\{T\}$ and $\mathbb{E}^{\boldsymbol{F}''}_{\tilde{q},\tilde{a}}\{T\}$.
Meanwhile, we have that policy $\boldsymbol{F}$ degenerates to policy $\boldsymbol{F}'$ and $\boldsymbol{F}''$ with $\epsilon$ as $1$ or $0$, respectively, where the corresponding $\epsilon'$ is equal to $1$ or $0$.

\vspace{-6mm}
\section{Proof of Theorem \ref{theorem_01}}
\label{Appendix_B}
\vspace{-5mm}

The main idea of the proof is to formulate the optimal policies for vertices of the optimal tradeoff curve $\mathcal{L}$ based on value iteration algorithm.
As presented in Fig. \ref{fig_weightedsum_piecewiselinear}, we obtain vertex $(\hat{P},\hat{D})$ as the only optimal power-delay pair of Lagrangian relaxation problem (\ref{eqn_lag_relax}) with the specific $\mu$.
The same optimal power-delay pair is obtained by the prime problem (\ref{eqn_stationary_optimization}) with $P_{\text{th}}=\hat{P}$.

To obtain optimal policies for vertices, we first formulate the MDP to minimize $D_{\boldsymbol{F}}+\mu{}P_{\boldsymbol{F}}$.
According to \cite[Theorem 9.1.8]{puterman2014markov}, we obtain the optimal deterministic policy $\boldsymbol{F}^\ast$ by using value iteration, which is presented in Algorithm \ref{algo_policy_iteration} with $\omega^{(m+1)}(q,a,s)$ \mbox{defined as}
{\setlength\abovedisplayskip {2pt plus 3pt minus 4pt}
	\setlength\belowdisplayskip {2pt plus 3pt minus 4pt}
	\vspace{-1mm}
\begin{equation}
	\omega^{(m+1)}(q,a,s)=\frac{1}{\alpha}q+\mu{}P(s)+\sum\nolimits_{a'=0}^{A}\gamma_{a,a'}\nu^{(m)}(q-s+a',a').
\label{def_h}
	\vspace{-2mm}
\end{equation}}%
Further, we have a unichain under the optimal policy $\boldsymbol{F}^\ast$ that is generated by Algorithm \ref{algo_policy_iteration}.

\setlength{\textfloatsep}{10pt}
\begin{algorithm}[t]
	\setstretch{0.85}
	\caption{Value Iteration Algorithm for Markov Decision Processes}
	\begin{algorithmic}[1]
		\State $m \gets 0$
		\ForAll{$q$ and $a$}
		\State $\nu^{(0)}(q,a) \gets$ arbitrary value // Initialization
		\EndFor
		\Repeat
		\ForAll{$q$ and $a$} // Policy Improvement:
		\State $s^{(m+1)}(q,a) \gets \arg\min_{s\in\mathcal{S}(q)}\{\omega^{(m+1)}(q,a,s)\}$
		\EndFor
		\ForAll{$q$ and $a$} // Policy Evaluation:
		\State $\nu^{(m+1)}(q,a) \gets \omega^{(m+1)}(q,a,s^{(m+1)}(q))$
		\EndFor
		\State $m \gets m+1$
		\Until{$s^{(m)}(q,a)=s^{(m-1)}(q,a)$ holds for all $q$ and $a$}
		\State $s^\ast(q,a) \gets s^{(m)}(q,a)$ for all $q$ and $a$
	\end{algorithmic}
	\label{algo_policy_iteration}
\end{algorithm}

Then, we show the threshold-based structure for policy $\boldsymbol{F}^\ast$.
Since the optimal policy is generated by an iteration process, we present the threshold-based structure by induction on $m$.
In particular, we first show the existence of thresholds for deterministic policy $\boldsymbol{F}^{(m+1)}=\{s^{(m+1)}(q,a):\forall{}q,a\}$ with the assumption that $\nu^{(m)}(q,a)$ is convex in $q$, i.e.,
{\setlength\abovedisplayskip {2pt plus 3pt minus 4pt}
\setlength\belowdisplayskip {2pt plus 3pt minus 4pt}
\vspace{-2mm}
\begin{equation}
	\nu^{(m)}(q-1,a)+\nu^{(m)}(q+1,a)\ge 2\nu^{(m)}(q,a).
\vspace{-2mm}
\end{equation}}%
To this end, we only need to show that transmission rate $s^{(m+1)}(q+1,a)$ is equal to $s^\ast$ or $s^\ast+1$

\noindent
when $s^{(m+1)}(q,a)$ is equal to $s^\ast$.
For a given arrival rate $a$, we then have that transmission rate $s^{(m+1)}(q,a)$ under deterministic policy $\boldsymbol{F}^{(m+1)}$ is monotone increasing on queue length $q$.
As a result, we have thresholds $q_{\boldsymbol{F}^{(m+1)}}(s,a)$ exist, and policy $\boldsymbol{F}^{(m+1)}$ satisfies Eq. (\ref{eqn_deterministic_threshold}).
With $s^\ast$ given as $\arg\min_{s\in\mathcal{S}(q)}\{\omega^{(m+1)}(q,a,s)\}$, we show the sufficient condition of thresholds' existence as
{\setlength\abovedisplayskip {1pt plus 2pt minus 3pt}
	\setlength\belowdisplayskip {1pt plus 2pt minus 3pt}
	\setlength\jot{-1pt}
\begin{align}
	\omega^{(m+1)}(q+1,a,s^\ast)&\le{}\omega^{(m+1)}(q+1,a,s^\ast-\delta)\label{h_1}\\
	\omega^{(m+1)}(q+1,a,s^\ast+1)&\le{}\omega^{(m+1)}(q+1,a,s^\ast+1+\delta)\label{h_2},
	\vspace{-1mm}
\end{align}}%
where $\delta\!\ge\!{}0$. 
Since $s^\ast$ minimizes $\omega^{(m+1)}(q,a,s)$ over $s\!\in\!\mathcal{S}(q)$, we rewrite Eqs. (\ref{h_1}) and (\ref{h_2}) as%
{\setlength\abovedisplayskip {1pt plus 2pt minus 3pt}
	\setlength\belowdisplayskip {1pt plus 2pt minus 3pt}
\begin{align}
	\omega^{(m+1)}(q,a,s^\ast\!-\!\delta)\!+\!\omega^{(m+1)}(q\!+\!1,a,s^\ast)&\le{}\omega^{(m+1)}(q,a,s^\ast)\!+\!\omega^{(m+1)}(q\!+\!1,a,s^\ast\!-\!\delta),\\
	\omega^{(m+1)}(q,a,s^\ast\!\!+\!\delta)\!+\!\omega^{(m+1)}(q\!+\!1,a,s^\ast\!\!+\!1)&\le{}\omega^{(m+1)}(q,a,s^\ast)\!+\!\omega^{(m+1)}(q\!+\!1,a,s^\ast\!+\!1\!+\!\delta),
\end{align}}%
respectively.
According to Eq. (\ref{def_h}), we can expand every components of the two inequalities. 
As a result, we immediately show the two inequalities following the convexity \mbox{of $P(s)$ and $\nu^{(m)}(q,a)$.}

We next show the convexity of $\nu^{(m+1)}(q,a)$ based on the threshold-based structure of $\boldsymbol{F}^{(m+1)}$.
In particular, the convexity of $\nu^{(m+1)}(q,a)$ is given from the definition of $\nu^{(m+1)}(q,a)$ as
\vspace{-1mm}
{\setlength\abovedisplayskip {2pt plus 6pt minus 8pt}
\setlength\belowdisplayskip {2pt plus 6pt minus 8pt}
\begin{equation}
	\omega^{(m+1)}(q+1,a,\tilde{s})+\omega^{(m+1)}(q-1,a,\breve{s})\ge{}2\omega^{(m+1)}(q,a,s^\ast),
\label{convex_v_eq}
\vspace{-1mm}
\end{equation}}%
where we have $\tilde{s}\!=\!\arg\min\limits_{s\in\mathcal{S}(q+1)}\{\omega^{(m+1)}(q\!+\!1,a,s)\}$ and $\breve{s}\!=\!\arg\min\limits_{s\in\mathcal{S}(q-1)}\{\omega^{(m+1)}(q\!-\!1,a,s)\}$.
Further, we have that $\tilde{s}$ and $s^\ast$ are selected from sets $\{s^\ast,s^\ast+1\}$ and $\{\breve{s},\breve{s}+1\}$, respectively.
We first present a sufficient condition for Eq. (\ref{convex_v_eq}) as
{\setlength\abovedisplayskip {2pt plus 6pt minus 8pt}
\setlength\belowdisplayskip {2pt plus 6pt minus 8pt}
\begin{equation}
	\omega^{(m+1)}(q+1,a,\tilde{s})+\omega^{(m+1)}(q-1,a,\breve{s})\ge{}\omega^{(m+1)}(q,a,s^\ast)+\omega^{(m+1)}(q,a,s'),
\label{suff_convex_v_eq}
\end{equation}}%
where $s'$ is an arbitrary transmission rate belonging to set $\mathcal{S}(q)$, and the sufficiency is guaranteed by $\omega^{(m+1)}(q,a,s^\ast)=\min_{s\in\mathcal{S}(q)}\{\omega^{(m+1)}(q,a,s)\}$.
Then, we show the sufficient condition by considering two cases, where $s^\ast$ is given as $\breve{s}$ or $\breve{s}+1$, respectively.
When $s^\ast=\breve{s}$, we set $s'=\tilde{s}$.
By expanding every components in Eq. (\ref{suff_convex_v_eq}), we verify the sufficient condition based on the convexity of $\nu^{(m)}(q,a)$.
When $s^\ast=\breve{s}+1$, we set $s'=\tilde{s}-1$, under which the sufficient condition holds based on the convexity of $P(s)$.
Since the initial $\nu^{(0)}(q,a)$ is convex in $q$, we have that deterministic policy $\boldsymbol{F}^{(m)}$ satisfies the threshold-based structure expressed in Eq. (\ref{eqn_deterministic_threshold}).

We finally supplement the proof for the degenerate case, in which one vertex may locate at a line segment generated by two vertices that adjacent with this vertex.
As a result, multiple points on the segment can minimize the Lagrangian relaxation problem.
In other words, we may not obtain the optimal policy for this vertex by Algorithm \ref{algo_policy_iteration}.
In this way, we present the optimal policy based on the sensitivity analysis of the equivalent LP problem.
With a slight drift in $P(s)$, we have that the degenerate case can be removed in the derived Lagrangian problem under the new $P(s)$ and the corresponding optimal policy will be unchanged.
Therefore, we can show that the optimal policy is threshold-based by using the same consideration as above.

\vspace{-4mm}
\section{Proof of Theorem \ref{theorem_last}}
\label{Appendix_D}
\vspace{-3mm}

Our proof starts with the observation that the optimal threshold-based policy can be presented as a convex combination of two adjacent deterministic threshold-based policies that correspond to two adjacent vertices on the optimal tradeoff curve $\mathcal{L}$. 
As a result, we shall only need to show Eq. (\ref{threshold_order}) in Theorem \ref{theorem_last} for the vertices of $\mathcal{L}$.
For each vertex of $\mathcal{L}$, we can also obtain the optimal policy for the system over a fading channel by using a value iteration, as shown in the proof of Theorem \ref{theorem_01}, through which we further show Eq. (\ref{threshold_order}) under condition in Eq. (\ref{power_condition}).
In particular, a sufficient condition of Eq. (\ref{threshold_order}) is given as 
{\setlength\abovedisplayskip {1pt plus 2pt minus 3pt}
	\setlength\belowdisplayskip {1pt plus 2pt minus 3pt}
	\setlength\jot{-1pt}
	\begin{equation}
	\omega^{(m+1)}(q,a,\iota^+,s^\ast)\!+\!\omega^{(m+1)}(q,a,\iota^-,s^\ast+\delta)\!\le\!{}\omega^{(m+1)}(q,a,\iota^+,s^\ast+\delta)\!+\!\omega^{(m+1)}(q,a,\iota^-,s^\ast),
	\vspace{-1mm}
	\end{equation}}%
\vspace{-2mm}%
where we denote the value function of the generalized system by $\omega^{(m+1)}(q,a,\iota,s)$, and define $s^\ast=\min_{s\in\mathcal{S}(q)}\{\omega^{(m+1)}(q,a,\iota,s)\}$.
Further, by expanding each component, we immediately show the sufficient condition under the condition in Eq. (\ref{power_condition}).
As a result, we have a greater transmission rate under the channel state $h_{\iota^+}$ than $h_{\iota^-}$.
With the threshold-based structure of the optimal policy, we finally show the order relation in Eq. (\ref{threshold_order}), which completes the proof.

\setstretch{1.15}
\footnotesize
\bibliographystyle{IEEEtran}
\bibliography{scheduling}

\begin{thebibliography}{10}
\providecommand{\url}[1]{#1}
\csname url@samestyle\endcsname
\providecommand{\newblock}{\relax}
\providecommand{\bibinfo}[2]{#2}
\providecommand{\BIBentrySTDinterwordspacing}{\spaceskip=0pt\relax}
\providecommand{\BIBentryALTinterwordstretchfactor}{4}
\providecommand{\BIBentryALTinterwordspacing}{\spaceskip=\fontdimen2\font plus
\BIBentryALTinterwordstretchfactor\fontdimen3\font minus
  \fontdimen4\font\relax}
\providecommand{\BIBforeignlanguage}[2]{{%
\expandafter\ifx\csname l@#1\endcsname\relax
\typeout{** WARNING: IEEEtran.bst: No hyphenation pattern has been}%
\typeout{** loaded for the language `#1'. Using the pattern for}%
\typeout{** the default language instead.}%
\else
\language=\csname l@#1\endcsname
\fi
#2}}
\providecommand{\BIBdecl}{\relax}
\BIBdecl

\bibitem{URLLC_req}
A.~Osseiran, F.~Boccardi, V.~Braun, K.~Kusume, P.~Marsch, M.~Maternia,
  O.~Queseth, M.~Schellmann, H.~Schotten, H.~Taoka, H.~Tullberg, M.~A.
  Uusitalo, B.~Timus, and M.~Fallgren, ``{Scenarios for 5G mobile and wireless
  communications: The vision of the METIS project},'' \emph{IEEE Communications
  Magazine}, vol.~52, no.~5, pp. 26--35, May 2014.

\bibitem{URLLC_req_3}
M.~Simsek, A.~Aijaz, M.~Dohler, J.~Sachs, and G.~Fettweis, ``{5G-enabled
  tactile internet},'' \emph{IEEE Journal on Selected Areas in Communications},
  vol.~34, no.~3, pp. 460--473, March 2016.

\bibitem{Power_efficient_1}
S.~{Buzzi}, C.~{I}, T.~E. {Klein}, H.~V. {Poor}, C.~{Yang}, and A.~{Zappone},
  ``{A survey of energy-efficient techniques for 5G networks and challenges
  ahead},'' \emph{IEEE Journal on Selected Areas in Communications}, vol.~34,
  no.~4, pp. 697--709, April 2016.

\bibitem{Power_efficient_2}
R.~Q. {Hu} and Y.~{Qian}, ``{An energy efficient and spectrum efficient
  wireless heterogeneous network framework for 5G systems},'' \emph{IEEE
  Communications Magazine}, vol.~52, no.~5, pp. 94--101, May 2014.

\bibitem{URLLC_allocation}
C.~She, C.~Yang, and T.~Q.~S. Quek, ``{Radio resource management for
  ultra-reliable and low-latency communications},'' \emph{IEEE Communications
  Magazine}, vol.~55, no.~6, pp. 72--78, June 2017.

\bibitem{cross_layer_delay_1}
Q.~Liu, S.~Zhou, and G.~B. {Giannakis}, ``{Cross-Layer combining of adaptive
  Modulation and coding with truncated ARQ over wireless links},'' \emph{IEEE
  Transactions on Wireless Communications}, vol.~3, no.~5, pp. 1746--1755, Sep.
  2004.

\bibitem{djonin2007mimo}
D.~V. Djonin and V.~Krishnamurthy, ``{MIMO transmission control in fading
  channels-a constrained Markov decision process formulation with monotone
  randomized policies},'' \emph{IEEE Transactions on Signal Processing},
  vol.~55, no.~10, pp. 5069--5083, 2007.

\bibitem{power_efficient}
A.~E. Gamal, C.~Nair, B.~Prabhakar, E.~Uysal-Biyikoglu, and S.~Zahedi,
  ``Energy-efficient scheduling of packet transmissions over wireless
  networks,'' in \emph{Proc. IEEE International Conference on Computer
  Communications (INFOCOM)}, June 2002, pp. 1773--1782.

\bibitem{power_ef_cross_layer_2}
U.~C. {Kozat}, I.~{Koutsopoulos}, and L.~{Tassiulas}, ``{A framework for
  cross-layer design of energy-efficient communication with QoS provisioning in
  multi-hop wireless networks},'' in \emph{Proc. IEEE International Conference
  on Computer Communications (INFOCOM)}, March 2004, pp. 1446--1456.

\bibitem{URLLC_cross_layer_3}
Z.~Hou, C.~She, Y.~Li, T.~Q.~S. Quek, and B.~Vucetic, ``{Burstiness aware
  bandwidth reservation for ultra-reliable and low-latency communications
  (URLLC) in tactile internet},'' \emph{IEEE Journal on Selected Areas in
  Communications}, vol.~36, no.~11, pp. 2401--2410, Nov. 2018.

\bibitem{power_ef_cross_layer}
J.~{Hu}, L.~{Yang}, and L.~{Hanzo}, ``Energy-efficient cross-layer design of
  wireless mesh networks for content sharing in online social networks,''
  \emph{IEEE Transactions on Vehicular Technology}, vol.~66, no.~9, pp.
  8495--8509, Sep. 2017.

\bibitem{collins1999transmission}
B.~Collins and R.~L. Cruz, ``Transmission policies for time varying channels
  with average delay constraints,'' in \emph{Proc. Allerton Conference on
  Communication, Control, and Computing (Allerton)}, 1999, pp. 709--717.

\bibitem{berry2002communication}
R.~A. Berry and R.~G. Gallager, ``Communication over fading channels with delay
  constraints,'' \emph{IEEE Transactions on Information Theory}, vol.~48,
  no.~5, pp. 1135--1149, 2002.

\bibitem{berry2013optimal}
R.~Berry, ``Optimal power-delay tradeoffs in fading channels--small-delay
  asymptotics,'' \emph{IEEE Transactions on Information Theory}, vol.~59,
  no.~6, pp. 3939--3952, June 2013.

\bibitem{rajan2004delay}
D.~Rajan, A.~Sabharwal, and B.~Aazhang, ``Delay-bounded packet scheduling of
  bursty traffic over wireless channels,'' \emph{IEEE Transactions on
  Information Theory}, vol.~50, no.~1, pp. 125--144, 2004.

\bibitem{chen2007optimal}
W.~Chen, Z.~Cao, and K.~B. Letaief, ``Optimal delay-power tradeoff in wireless
  transmission with fixed modulation,'' in \emph{Proc. IEEE International
  Workshop on Cross Layer Design (IWCLD)}, 2007, pp. 60--64.

\bibitem{goyal2003power}
M.~Goyal, A.~Kumar, and V.~Sharma, ``Power constrained and delay optimal
  policies for scheduling transmission over a fading channel,'' in \emph{Proc.
  IEEE International Conference on Computer Communications (INFOCOM)}, 2003,
  pp. 311--320.

\bibitem{ata2005dynamic}
B.~Ata, ``Dynamic power control in a wireless static channel subject to a
  quality-of-service constraint,'' \emph{Operations Research}, vol.~53, no.~5,
  pp. 842--851, 2005.

\bibitem{ngo2010monotonicity}
M.~H. Ngo and V.~Krishnamurthy, ``{Monotonicity of constrained optimal
  transmission policies in correlated fading channels with ARQ},'' \emph{IEEE
  Transactions on Signal Processing}, vol.~58, no.~1, pp. 438--451, 2010.

\bibitem{MDP_communication_1}
L.~{Liu}, A.~{Chattopadhyay}, and U.~{Mitra}, ``{On solving MDPs with large
  state space: Exploitation of policy structures and spectral properties},''
  \emph{IEEE Transactions on Communications}, Early Access, 2019.

\bibitem{MDP_communication_2}
\BIBentryALTinterwordspacing
N.~Sharma, N.~Mastronarde, and J.~Chakareski, ``Accelerated structure-aware
  reinforcement learning for delay-sensitive energy harvesting wireless
  sensors,'' \emph{CoRR}, vol. abs/1807.08315, 2018. [Online]. Available:
  \url{http://arxiv.org/abs/1807.08315}
\BIBentrySTDinterwordspacing

\bibitem{chen2017delaytcom}
X.~Chen, W.~Chen, J.~Lee, and N.~B. Shroff, ``{Delay-optimal buffer-aware
  scheduling with adaptive transmission},'' \emph{IEEE Transactions on
  Communications}, vol.~65, no.~7, pp. 2917--2930, July 2017.

\bibitem{MwangTCOM}
M.~{Wang}, J.~{Liu}, W.~{Chen}, and A.~{Ephremides}, ``Joint queue-aware and
  channel-aware delay optimal scheduling of arbitrarily bursty traffic over
  multi-state time-varying channels,'' \emph{IEEE Transactions on
  Communications}, vol.~67, no.~1, pp. 503--517, Jan 2019.

\bibitem{ARQ}
J.~Liu, W.~Chen, and K.~B. Letaief, ``{Delay optimal scheduling for ARQ-aided
  power-constrained packet transmission over multi-state fading channels},''
  \emph{IEEE Transactions on Wireless Communications}, vol.~16, no.~11, pp.
  7123--7137, Nov. 2017.

\bibitem{xiang_ICC_2017}
X.~Chen, W.~Chen, J.~Lee, and N.~B. Shroff, ``Delay-optimal probabilistic
  scheduling in green communications with arbitrary arrival and adaptive
  transmission,'' in \emph{Proc. IEEE International Conference on
  Communications (ICC)}, May 2017, pp. 1--6.

\bibitem{self_similar_2}
V.~{Paxson} and S.~{Floyd}, ``Wide area traffic: The failure of poisson
  modeling,'' \emph{IEEE/ACM Transactions on Networking}, vol.~3, no.~3, pp.
  226--244, June 1995.

\bibitem{puterman2014markov}
M.~L. Puterman, \emph{Markov decision processes: Discrete stochastic dynamic
  programming}.\hskip 1em plus 0.5em minus 0.4em\relax John Wiley \& Sons,
  2014.

\bibitem{kao1997introduction}
E.~P. Kao, \emph{{An introduction to stochastic processes}}.\hskip 1em plus
  0.5em minus 0.4em\relax Cengage Learning, 1997.

\bibitem{altman1999constrained}
E.~Altman, \emph{Constrained Markov decision processes}.\hskip 1em plus 0.5em
  minus 0.4em\relax CRC Press, 1999.

\end{thebibliography}
\normalsize

\end{document}